\documentclass[final,times,twocolumn,sort&compress]{elsarticle}
\usepackage{hyperref}
\usepackage{color}
\usepackage{textcomp}
\usepackage{amsmath}
\usepackage{amssymb}
\usepackage{graphicx}
\usepackage{esint}
\usepackage{dcolumn}
\usepackage{gensymb}

\journal{Journal of Magnetism and Magnetic Materials}

\begin{document}

\begin{frontmatter}

\title{Heterobimetallic Dy-Cu coordination compound as a classical-quantum ferrimagnetic chain of regularly alternating Ising and Heisenberg spins\tnoteref{grants}}
\tnotetext[grants]{This work was supported by the Brazilian grant agencies FAPEAL (Alagoas State Research agency), CNPq, CAPES, FAPEMIG and by the Slovak grant agencies under the contract Nos. VEGA 1/0331/15 and APVV-14-0073.}

\author[a1]{J. Torrico\corref{mycorrespondingauthor}}\cortext[mycorrespondingauthor]{Corresponding author}\ead{jordanatorrico@gmail.com}
\author[a2]{J. Stre\v{c}ka}
\author[a3]{M. Hagiwara}
\author[a4]{O. Rojas}
\author[a4]{S. M. de Souza}
\author[a5]{Y. Han}
\author[a6]{Z. Honda}
\author[a1]{M. L. Lyra}

\address[a1]{Instituto de F\'isica, Universidade Federal de Alagoas, 57072-970, Macei\'o-AL, Brazil}
\address[a2]{Institute of Physics, Faculty of Science, P. J. \v{S}af\'arik University, Park Angelinum 9, 040 01 Ko\v{s}ice, Slovak Republic}
\address[a3]{Center for Advanced High Magnetic Field Science,Graduate School of Science, Osaka University,1-1 Machikaneyama, Toyonaka, Osaka 560-0043, Japan}
\address[a4]{Departamento de F\'isica, Universidade Federal de Lavras, 37200-000, Lavras-MG, Brazil}
\address[a5]{Wuhan National High Magnetic Field Center, Huazhong University of Science and Technology, Wuhan 430074, China}
\address[a6]{Graduate School of Science and Engineering, Saitama University, 255 Shimo-Okubo, Sakura-ku, Saitama 338-8570, Japan}

\begin{abstract}
A classical-quantum chain composed of regularly alternating Ising and Heisenberg spins is rigorously solved by considering two distinct local anisotropy axes of the Ising spins. The ground-state phase diagram and magnetization curves are examined depending on a spatial orientation of the applied magnetic field. The phase diagram totally consists of four distinct phases and a few macroscopically degenerate points, where an outstanding coexistence of perfect order and complete disorder occurs within the so-called 'half-fire, half-ice' state. The zero-temperature magnetization curves generally exhibit a smooth dependence on a magnetic field owing to a canting angle between two coplanar anisotropy axes of the Ising spins, which enforces a misalignment of the magnetization vector from a direction of the applied magnetic field. It is evidenced that the investigated spin-chain model reproduces magnetic features of the heterobimetallic coordination compound Dy(NO)$_{3}$(DMSO)$_{2}$Cu(opba)(DMSO)$_{2}$ (DMSO=dimethylsulfoxide, opba=orthophenylenebisoxamato). The high-field magnetization data reported for the powder sample of this polymeric coordination compound generally display a substantial smoothing on account of a powder averaging.
\end{abstract}


\end{frontmatter}

\section{Introduction}

Magnetization curves of one-dimensional (1D) quantum spin chains generally exhibit several intriguing features of purely quantum origin, which are most evidently manifested at zero temperature because a spurious effect of thermal fluctuations is completely suppressed \cite{hon04,mik04,tak11}. Interestingly, several polymeric coordination compounds can be truly classified from the magnetic viewpoint as experimental representatives of 1D quantum spin chains \cite{jon74,mil02}. From the theoretical point of view, the 1D quantum spin chains provide important cornerstones of the quantum theory of magnetism, since they belong to valuable examples of exactly solved lattice-statistical models unaffected by crude approximations \cite{mat93}. The spin-$1/2$ Ising-Heisenberg chain \cite{lie61} with regularly alternating Ising and Heisenberg bonds is historically the first exactly solved spin model, which displays a remarkable quantum phase transition driven by external parameters \cite{yao02,str10,liu15}. It should be nevertheless mentioned that the Ising-Heisenberg spin chains were for a long time regarded only as a mathematical curiosity without any close resemblance to real-world magnetic compounds.

\begin{figure}[t]
\includegraphics[width=0.5\textwidth]{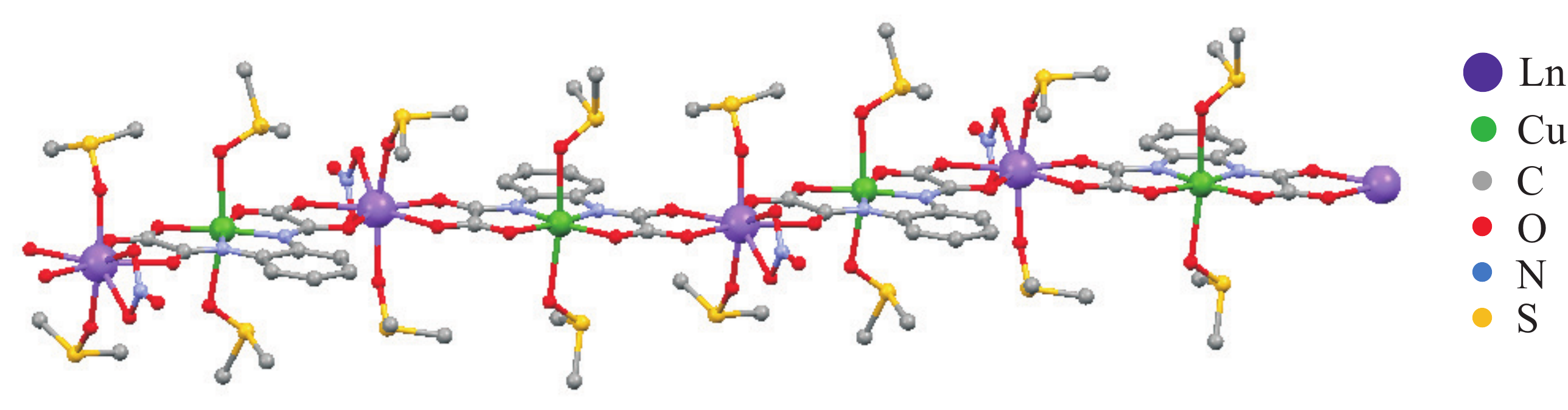}
\vspace{-0.7cm}
\caption{(Color online) A part of the crystal structure of the 3d-4f bimetallic coordination polymer Ln(NO$_3$)(DMSO)$_2$Cu(opba)(DMSO)$_2$ (Ln = Gd--Er) visualized by adapting the crystallographic data deposited at The Cambridge Crystallographic Data Centre according to Ref. \cite{cal08}. Coloring scheme for the atom labelling: Ln (violet), Cu (green), C (grey), O (red), N (blue), S (yellow).}
\label{dycu}
\end{figure}

Recently, it has been verified that a few exactly solved Ising-Heisenberg models may capture basic magnetic features of some polymeric coordination compounds such as Cu(3-Clpy)$_2$(N$_3$)$_2$ \cite{str05}, [(CuL)$_2$Dy][Mo(CN)$_8$] \cite{heu10,bel14} and [Fe(H$_2$O)(L)][Nb(CN)$_8$][Fe(L)] \cite{sah12}. The heterobimetallic polymeric complex Dy(NO$_3$)(DMSO)$_2$Cu(opba)(DMSO)$_2$, which will be hereafter referred to as Dy-Cu, involves 1D chain of exchange-coupled Dy$^{3+}$ and Cu$^{2+}$ ions as a magnetic backbone \cite{cal08} (see Fig. \ref{dycu}). Consequently, the polymeric compound Dy-Cu can be regarded as an experimental realization of the spin-$1/2$ Ising-Heisenberg chain with regularly alternating Ising and Heisenberg spins, which capture a magnetic behavior of highly anisotropic Dy$^{3+}$ and almost isotropic Cu$^{2+}$ magnetic ions, respectively \cite{str12,han13}. This compound is quite interesting also from the perspective of rise of pairwise thermal entanglement as exemplified in Ref. \cite{roj14}. A closer inspection of crystallographic data \cite{cal08} however reveals that there exist two inequivalent orientations of coordination polyhedra of Dy$^{3+}$ ions within the Dy-Cu polymeric chain, which have not been taken into consideration in previous studies \cite{str12,han13,roj14}. Bearing this in mind, we will investigate in the present work the spin-$1/2$ Ising-Heisenberg chain with regularly alternating Ising and Heisenberg spins in an arbitrarily oriented magnetic field by considering two different local anisotropy axes of the Ising spins resembling two crystallographically inequivalent positions of Dy$^{3+}$ ions within the Dy-Cu polymeric chain. In particular, our attention will be focused on a role of the canting of local anisotropy axes on the ground state and magnetization process.

The organization of this paper is as follows. The model and its exact solution is presented in Section \ref{model}. The most interesting results for the ground-states phase diagrams, the magnetization curves and susceptibility are comprehensively discussed in Section \ref{result}. The high-field magnetization data recorded on the Dy-Cu compound are confronted with the respective theoretical predictions in Section \ref{expthe}. Finally, the paper ends up with several concluding remarks given in Section \ref{conclusion}.

\section{The model and its solution}
\label{model}

\begin{figure}[t]
\includegraphics[scale=0.53]{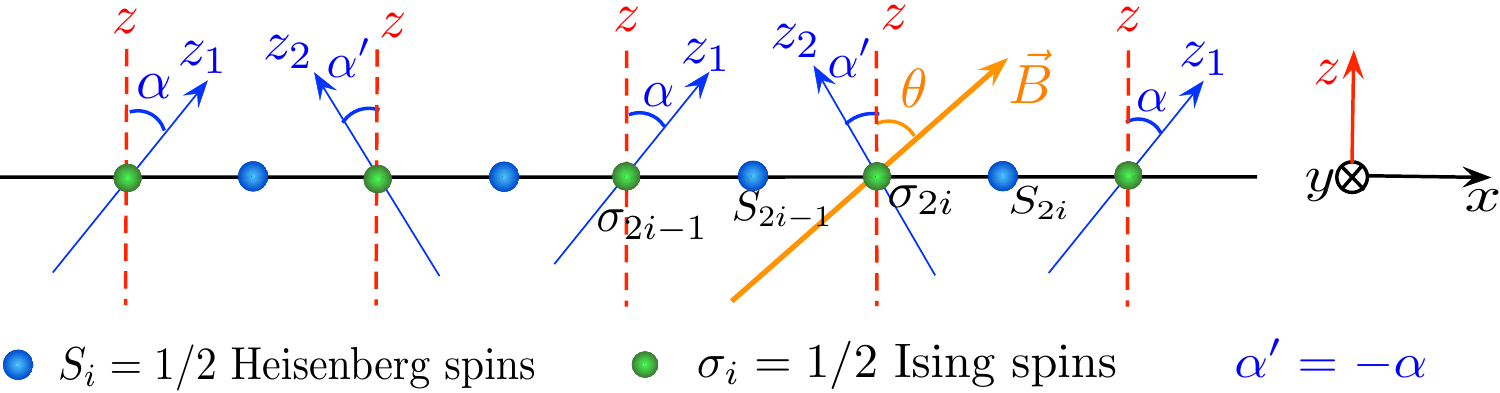} 
\caption{\label{fig:1} (Color online) A schematic representation of 1D chain with regularly alternating Ising and Heisenberg spins. The angle $\alpha$ (-$\alpha$) determines the canting of the local anisotropy axis $z_1$ ($z_2$) from the global frame $z$-axis for odd (even) Ising spins $\sigma_{2i-1}^{z_{1}}$ ($\sigma_{2i}^{z_{2}}$) so that $2 \alpha$ is the canting angle between two coplanar anisotropy axes. The angle $\theta$ determines the tilting of the magnetic field from the global frame $z$-axis.}
\end{figure}

Let us consider the spin-1/2 Ising-Heisenberg chain, in which the Ising spins $\sigma=1/2$ with two different co-planar anisotropy axes $z_1$ and $z_2$ regularly alternate with the Heisenberg spins $S=1/2$ as it is schematically depicted in Fig.~\ref{fig:1}. The local anisotropy axis $z_1$ ($z_2$) of the Ising spins $\sigma_{2i-1}^{z_{1}}$ ($\sigma_{2i}^{z_{2}}$) 
on odd (even) lattice positions is canted by the angle $\alpha$ (-$\alpha$) from the global frame $z$-axis, which means that the angle $2\alpha$ determines the overall canting between two coplanar anisotropy axes $z_1$ and $z_2$. The Heisenberg spins are coupled to their nearest-neighbor Ising spins through the antiferromagnetic coupling $J<0$ projected into the respective anisotropy axes, whereas the relevant projections of the spin operators into the anisotropy axes $z_1$ and $z_2$ are denoted as $S_{i}^{z_{1}}$ and $S_{i}^{z_{2}}$, respectively. Moreover, we will consider the effect of the external magnetic field $B$, whose spatial orientation with respect to the global frame $z$-axis is given by the azimuthal angle $\theta$ (see Fig.~\ref{fig:1}). The spin-1/2 Ising-Heisenberg chain with regularly alternating Ising and Heisenberg spins can be subsequently defined through the Hamiltonian
\begin{align}
\mathcal{H}= & -J\overset{N/2}{\underset{i=1}{\sum}}[\sigma_{2i-1}^{z_{1}}(S_{2i-2}^{z_{1}}
+S_{2i-1}^{z_{1}})+\sigma_{2i}^{z_{2}}(S_{2i}^{z_{2}}+S_{2i-1}^{z_{2}})]\nonumber \\
& -h^{z_{1}} \! \sum_{i=1}^{N/2}\! \sigma_{2i-1}^{z_{1}} - h^{z_{2}} \! \sum_{i=1}^{N/2}\! \sigma_{2i}^{z_{2}}
  -h^z \! \sum_{i=1}^{N}\! S_{i}^{z} - h^x \! \sum_{i=1}^{N}\! S_{i}^{x}\!,
\label{eq:1}
\end{align}
where $h^{z_{1}} = g_{1}^{z_{1}} \mu_{\rm B} B $cos($\alpha-\theta)$ and $h^{z_{2}} = g_{1}^{z_{2}} \mu_{\rm B} B $cos($\alpha+\theta)$ determine  projections of the external magnetic field $B$ towards the local anisotropy axes $z_1$ and $z_2$ of the Ising spins $\sigma_{2i-1}^{z_{1}}$ and $\sigma_{2i}^{z_{2}}$ on odd and even lattice positions, respectively, $g_{1}^{z_{1}}$ and $g_{1}^{z_{2}}$ are the respective g-factors of the Ising spins and $\mu_{\rm B}$ is the Bohr magneton. Similarly, $h^{z}= g_2^z \mu_{\rm B} B \cos \theta$ and $h^{x}= g_2^x \mu_{\rm B} B \sin \theta$ determine two orthogonal projections of the external magnetic field acting on the Heisenberg spins, whereas $g_2^z$ and $g_2^x$ are the respective spatial components of the g-factor of the Heisenberg spins.

The Hamiltonian (\ref{eq:1}) can be alternatively rewritten as a sum of the cell Hamiltonians
\begin{equation}
\mathcal{H}=\sum_{i=1}^{N/2}\left(\mathcal{H}_{2i-1}+\mathcal{H}_{2i}\right),
\label{eq:2}
\end{equation}
which involve all interaction and field terms related to one Heisenberg spin at odd or even lattice position
\begin{align}
\mathcal{H}_{2i-1}=& -\frac{h^{z_{1}}}{2} \sigma_{2i-1}^{z_{1}} - \frac{h^{z_{2}}}{2} \sigma_{2i}^{z_{2}}
                     - h_{2i-1}^{z} S_{2i-1}^{z} - h_{2i-1}^{x} S_{2i-1}^{x}, \nonumber \\
\mathcal{H}_{2i}=& -\frac{h^{z_{2}}}{2} \sigma_{2i}^{z_{2}} - \frac{h^{z_{1}}}{2} \sigma_{2i+1}^{z_{1}} - h_{2i}^{z} S_{2i}^{z} - h_{2i}^{x} S_{2i}^{x}. \label{eq:3}
\end{align}
For abbreviation purposes, we have introduced in above the following notation for effective longitudinal and transverse fields acting on the Heisenberg spins
\begin{align}
  h_{2i-1}^{z} &=J\cos\alpha\left(\sigma_{2i-1}^{z_{1}}+\sigma_{2i}^{z_{2}}\right)+g_{2}^{z}\mu_{\rm B} B \cos \theta, \nonumber\\
  h_{2i-1}^{x}&=J\sin\alpha\left(\sigma_{2i-1}^{z_{1}}-\sigma_{2i}^{z_{2}}\right)+g_{2}^{x}\mu_{\rm B} B \sin \theta,\nonumber\\
  h_{2i}^{z}&=J\cos\alpha\left(\sigma_{2i}^{z_{2}}+\sigma_{2i+1}^{z_{1}}\right)+g_{2}^{z}\mu_{\rm B} B \cos \theta, \nonumber\\
  h_{2i}^{x}&=-J\sin\alpha\left(\sigma_{2i}^{z_{2}}-\sigma_{2i+1}^{z_{1}}\right)+g_{2}^{x}\mu_{\rm B} B \sin \theta.
\label{eq:hef}
\end{align}
It is quite obvious that the cell Hamiltonians (\ref{eq:3}) commute with each other, i.e. $[\mathcal{H}_{i},\mathcal{H}_{j}] = 0$, which means that they belong to orthogonal Hilbert subspaces and can be diagonalized independently of each other. The cell Hamiltonians (\ref{eq:3}) can be straightforwardly diagonalized by a local spin-rotation transformation, which gives the following two eigenvalues of the odd cell Hamiltonian $\mathcal{H}_{2i-1}$
\begin{align}
\epsilon_{2i-1}^{\pm}= -\frac{h^{z_{1}}}{2} \sigma_{2i-1}^{z_{1}} - \frac{h^{z_{2}}}{2} \sigma_{2i}^{z_{2}} \pm\frac{1}{2}\sqrt{\left(h_{2i-1}^{z}\right)^{2}+\left(h_{2i-1}^{x}\right)^{2}}
\label{eq:5}
\end{align}
for the corresponding eigenvectors
\begin{eqnarray}
\left|\psi^{\pm}\right\rangle_{2i-1}  =  \frac{1}{\sqrt{(a_{2i-1}^{\pm})^{2}+1}}\left(-a_{2i-1}^{\pm}\left|\uparrow\right\rangle_{2i-1}+\left|\downarrow\right\rangle_{2i-1}\right).
\label{eq:6}
\end{eqnarray}
Here, $a_{i}^{\pm}=h_{i}^{x}/[h_{i}^{z}\pm\sqrt{(h_{i}^{z})^{2}+(h_{i}^{x})^{2}}]$ determines probability amplitudes for two available states of the Heisenberg spin from the $i$-th lattice position. Similarly, two eigenvalues of the even cell Hamiltonian $\mathcal{H}_{2i}$ read
\begin{eqnarray}
\epsilon_{2i}^{\pm}=-\frac{h^{z_{2}}}{2} \sigma_{2i}^{z_{2}} - \frac{h^{z_{1}}}{2} \sigma_{2i+1}^{z_{1}} \pm\frac{1}{2}\sqrt{\left(h_{2i}^{z}\right)^{2}+\left(h_{2i}^{x}\right)^{2}},
\label{eq:7}
\end{eqnarray}
whereas the corresponding eigenvectors are given by
\begin{eqnarray}
\left|\psi^{\pm}\right\rangle_{2i} =  \frac{1}{\sqrt{(a_{2i}^{\pm})^{2}+1}}\left(-a_{2i}^{\pm}\left|\uparrow\right\rangle_{2i} +\left|\downarrow\right\rangle_{2i}\right).
\label{eq:8}
\end{eqnarray}

On account of a regular alternation of two different local anisotropy axes of the Ising spins the elementary unit cell of the spin-1/2 Ising-Heisenberg chain incorporates two Ising and two Heisenberg spins. Hence, it follows that the ground state of the spin-1/2 Ising-Heisenberg chain can be built up from the lowest-energy eigenstate of the unit cell with two Ising and two Heisenberg spins. The overall energy spectrum of the unit cell of the spin-1/2 Ising-Heisenberg chain consists of four eigenvalues
\begin{align}
\varepsilon_{0} & = \epsilon_{2i-1}^{-}+\epsilon_{2i}^{-},\label{eq:9}\\
\varepsilon_{1} & = \epsilon_{2i-1}^{-}+\epsilon_{2i}^{+},\label{eq:10}\\
\varepsilon_{2} & =  \epsilon_{2i-1}^{+}+\epsilon_{2i}^{-},\label{eq:11}\\
\varepsilon_{3} & =  \epsilon_{2i-1}^{+}+\epsilon_{2i}^{+},\label{eq:12}
\end{align}
which still depend through Eqs. (\ref{eq:hef}), (\ref{eq:5}) and (\ref{eq:7}) on the particular spin orientations of the Ising spins. The eigenvalues (\ref{eq:9})-(\ref{eq:12}) and their corresponding eigenvectors will be subsequently used in order to determine the ground-state phase diagram depending on a canting angle between the anisotropy axes of the Ising spins and a spatial orientation of the external magnetic field.

The exact solution for the investigated spin system can be obtained by making use of the transfer-matrix method \cite{bax82}, which allows the following factorization of the partition function
\begin{eqnarray}
\mathcal{Z}\!\!\!&=&\!\!\!\sum_{\{\sigma\}}\prod_{i=1}^{N/2} \left[\rm{Tr}_{S_{2i-1}}e^{-\beta \mathcal{H}_{2i-1}}\right]
\left[\rm{Tr}_{S_{2i}}e^{-\beta \mathcal{H}_{2i}}\right],\nonumber\\
\!\!\!&=&\!\!\!\sum_{\{\sigma\}}\prod_{i=1}^{N/2}\mathbf{V}_1(\sigma_{2i-1}^{z_1},\sigma_{2i}^{z_2})\mathbf{V}_2(\sigma_{2i}^{z_2},\sigma_{2i+1}^{z_1}).
\label{eq:14}
\end{eqnarray}
Here, $\beta=1/(k_B T)$, $k_B$ is the Boltzmann's constant, $T$ is the absolute temperature and the summation $\sum_{\{\sigma\}}$ runs over all possible spin states of the Ising spins. The expressions $\mathbf{V}_1(\sigma_{2i-1}^{z_1},\sigma_{2i}^{z_2})$ and $\mathbf{V}_2(\sigma_{2i}^{z_2},\sigma_{2i+1}^{z_1})$ can be viewed as transfers matrices associated with the local Hamiltonians (\ref{eq:3}), which  involve all interactions terms of a single Heisenberg spin placed either on odd or even lattice position. The transfer matrices $\mathbf{V}_1(\sigma_{2i-1}^{z_1},\sigma_{2i}^{z_2})$ and $\mathbf{V}_2(\sigma_{2i}^{z_2},\sigma_{2i+1}^{z_1})$ are then defined as
\begin{align}
\mathbf{V}_1&=
\left( \begin{array}{cccc}
V_1\left(\frac{1}{2},\frac{1}{2}\right) & V_1\left(\frac{1}{2},-\frac{1}{2}\right)\\
V_1\left(-\frac{1}{2},\frac{1}{2}\right) & V_1\left(-\frac{1}{2},-\frac{1}{2}\right)\\
\end{array}\right)
\label{eq:15},\\
\mathbf{V}_2&=
\left( \begin{array}{cccc}
V_2\left(\frac{1}{2},\frac{1}{2}\right) & V_2\left(\frac{1}{2},-\frac{1}{2}\right)\\
V_2\left(-\frac{1}{2},\frac{1}{2}\right) & V_2\left(-\frac{1}{2},-\frac{1}{2}\right)\\
\end{array}\right)
\label{eq:16}.
\end{align}
It can be easily checked that the transfer matrices $\mathbf{V}_1$ and $\mathbf{V}_2$ are mutually transposed with respect to each other ($\mathbf{V}_1 = \mathbf{V}_2^T$), which means that they both can be defined just through four different matrix elements
\begin{align*}
V_{++} &\equiv V_1\left(\frac{1}{2},\frac{1}{2}\right)=V_2\left(\frac{1}{2},\frac{1}{2}\right),\\
V_{+-} &\equiv V_1\left(\frac{1}{2},-\frac{1}{2}\right)=V_2\left(-\frac{1}{2},\frac{1}{2}\right),\\
V_{-+} &\equiv V_1\left(-\frac{1}{2},\frac{1}{2}\right)=V_2\left(\frac{1}{2},-\frac{1}{2}\right),\\
V_{--} &\equiv V_1\left(-\frac{1}{2},-\frac{1}{2}\right)=V_2\left(-\frac{1}{2},-\frac{1}{2}\right).
\end{align*}
Further, the product of both transfer matrices $\mathbf{V}_1 \mathbf{V}_2$ can be viewed as the transfer matrix of the unit cell given by
\begin{eqnarray}
\mathbf{V}_1\mathbf{V}_2=
\left( \begin{array}{cccc}
V_{++}^2+V_{+-}^2 & V_{++}V_{-+}+V_{+-}V_{--}\\
V_{++}V_{-+}+V_{+-}V_{--} & V_{-+}^2+V_{--}^2\\
\end{array}\right),
\label{eq:17}
\end{eqnarray}
the diagonalization of which allows a straightforward calculation of the partition function according to
\begin{eqnarray}
\mathcal{Z}=\rm{Tr}\left(\mathbf{V}_1\mathbf{V}_2\right)^{\frac{N}{2}}=\lambda_1^{\frac{N}{2}}+\lambda_2^{\frac{N}{2}}.
\label{eq:18}
\end{eqnarray}
The respective eigenvalues of the transfer matrix $\mathbf{V}_1\mathbf{V}_2$ can be expressed as
\begin{eqnarray}
\lambda_{\pm}=\frac{1}{2} \left[\rm{Tr}(\mathbf{V}_1\mathbf{V}_2)\pm\sqrt{\rm{Tr}(\mathbf{V}_1\mathbf{V}_2)^2-\rm{Det}(\mathbf{V}_1\mathbf{V}_2)}\right].
\label{eq:19}
\end{eqnarray}
For completeness, let us quote the explicit form of the transfer-matrix elements
\begin{eqnarray}
V_{++}\!\!\!&=&\!\!\!2\exp \left[\frac{\beta}{2}(h_{2i-1}^{z_1}+h_{2i}^{z_2})\right]\nonumber\\
&~&\times\cosh\left\{\frac{\beta}{2}\left[\frac{J^2}{2}(1+\cos2\alpha)+q_{+}\right]^{\frac{1}{2}}\right\}, \label{eq:20}\\
V_{+-}\!\!\!&=&\!\!\!2\exp \left[\frac{\beta}{2}(h_{2i-1}^{z_1}-h_{2i}^{z_2})\right]\nonumber\\
&~&\times\cosh\left\{\frac{\beta}{2}\left[\frac{J^2}{2}(1-\cos2\alpha)+p_{+}\right]^{\frac{1}{2}}\right\}, \label{eq:21}\\
V_{-+}\!\!\!&=&\!\!\!2\exp \left[-\frac{\beta}{2}(h_{2i-1}^{z_1}-h_{2i}^{z_2})\right]\nonumber\\
&~&\times\cosh\left\{\frac{\beta}{2}\left[\frac{J^2}{2}(1-\cos2\alpha)+p_{-}\right]^{\frac{1}{2}}\right\}, \label{eq:22}\\
V_{--}\!\!\!&=&\!\!\!2\exp \left[-\frac{\beta}{2}(h_{2i-1}^{z_1}+h_{2i}^{z_2})\right]\nonumber\\
&~&\times\cosh\left\{\frac{\beta}{2}\left[\frac{J^2}{2}(1+\cos2\alpha)+q_{-}\right]^{\frac{1}{2}}\right\}, \label{eq:23}
\end{eqnarray}
where
\begin{eqnarray*}
q_{\pm}=(g_2^z\mu_BB^z)^2+(g_2^x\mu_BB^x)^2\pm2Jg_2^z\mu_BB^z\cos\alpha, \\
p_{\pm}=(g_2^z\mu_BB^z)^2+(g_2^x\mu_BB^x)^2\pm2Jg_2^x\mu_BB^x\sin\alpha.
\end{eqnarray*}
In thermodynamic limit $N \to \infty$, the Helmholtz free energy per block can be expressed in terms of the larger transfer-matrix eigenvalue
\begin{eqnarray}
f=-k_BT\lim_{N\rightarrow\infty}\frac{1}{N}\ln\mathcal{Z}=-\frac{1}{2}k_BT\ln\lambda_{+}. \\ \nonumber
\label{eq:24}
\end{eqnarray}
The Helmholtz free energy (\ref{eq:24}) subsequently allows a straightforward calculation of the magnetization and other thermodynamic quantities of interest. The total magnetization per unit cell can be for instance calculated from the relation
\begin{eqnarray}
m_t=-\frac{\partial f}{\partial B}.
\label{eq:25}
\end{eqnarray}
To get a deeper insight into the magnetization process it is also of particular interest to calculate the mean values of longitudinal and transverse  components of the Heisenberg spins
\begin{align}
m_2^z &\equiv g_2^z \mu_{\rm B} \langle S_i^z \rangle = -g_2^z \mu_{\rm B} \frac{\partial f}{\partial h^z}, \label{eq:26} \\
m_2^x &\equiv g_2^x \mu_{\rm B} \langle S_i^x \rangle = -g_2^x \mu_{\rm B} \frac{\partial f}{\partial h^x}, \label{eq:27}
\end{align}
as well as the mean values of the projections of the Ising spins towards their local easy axis
\begin{align}
m_1^{z_1} &\equiv g_1^{z_1} \mu_{\rm B} \langle \sigma_{2i-1}^{z_1} \rangle =-g_1^{z_1} \mu_{\rm B}\frac{\partial f}{\partial h^{z_1}}, \label{eq:28}\\
m_1^{z_2} &\equiv g_1^{z_2} \mu_{\rm B}\langle \sigma_{2i}^{z_2} \rangle =-g_1^{z_2} \mu_{\rm B}\frac{\partial f}{\partial h^{z_2}}. \label{eq:29}
\end{align}
The powder magnetization represents another quantity of particular interest, because it enables characterization of the magnetization process for  polycrystalline systems as usually reported in experimental studies \cite{cal08,str12,han13}. The powder magnetization can be calculated from the total magnetization (\ref{eq:25}) by performing the powder averaging, which consists in integrating the total magnetization over all possible relative orientations of the magnetic field
\begin{equation}
m_{\mathrm{powder}}=\intop_{0}^{\frac{\pi}{2}}m_t(\theta)\sin(\theta)d\theta.\label{eq:30}
\end{equation}
where
\begin{align}
m_t(\theta)=& m_1^{z_1}\cos(\alpha-\theta)+ m_1^{z_2}\cos(\alpha+\theta)\nonumber\\
&+m_2^{x}\sin(\theta)+m_2^{z}\cos(\theta).
\label{eq:31}
\end{align}

The isothermal susceptibility can also be easily calculated from the total magnetization (\ref{eq:25}) according to the formula
\begin{eqnarray}
\chi_t=\frac{\partial m_t}{\partial B},
\label{eq:34}
\end{eqnarray}
whereas its longitudinal ($\chi_z$) and transverse ($\chi_x$) components follow from
\begin{eqnarray}
\chi_z&=&\frac{\partial m_t}{\partial (B \cos \theta)}, \\
\label{eq:32}
\chi_x&=&\frac{\partial m_t}{\partial (B \sin \theta)}.
\label{eq:33}
\end{eqnarray}
The powder susceptibility can be then approximated by a simple average over the contributions along three orthogonal axes
\begin{eqnarray}
\chi_{powder}\cong\frac{2\chi_x+\chi_z}{3}.
\label{eq:35}
\end{eqnarray}

\section{Results and discussion}
\label{result}

Let us proceed to a discussion of the most interesting results for the investigated spin-1/2 Ising-Heisenberg chain with the antiferromagnetic nearest-neighbor exchange coupling $J<0$, the absolute value of which will henceforth serve as an energy unit. To reduce the number of free parameters, we will further consider a unique value of g-factor $g_{1}^{z_{1}}=g_{1}^{z_{2}}=20$ of the Ising spins and the isotropic value of g-factor of the Heisenberg spins $g_2^x = g_2^z = 2$, which nearly coincide with usual values of gyromagnetic ratio for Dy$^{3+}$ and Cu$^{2+}$ ions creating a magnetic backbone of the DyCu polymeric chain \cite{cal08}.

\subsection{Ground-state phase diagram}

\begin{figure}
\includegraphics[width=0.5\textwidth]{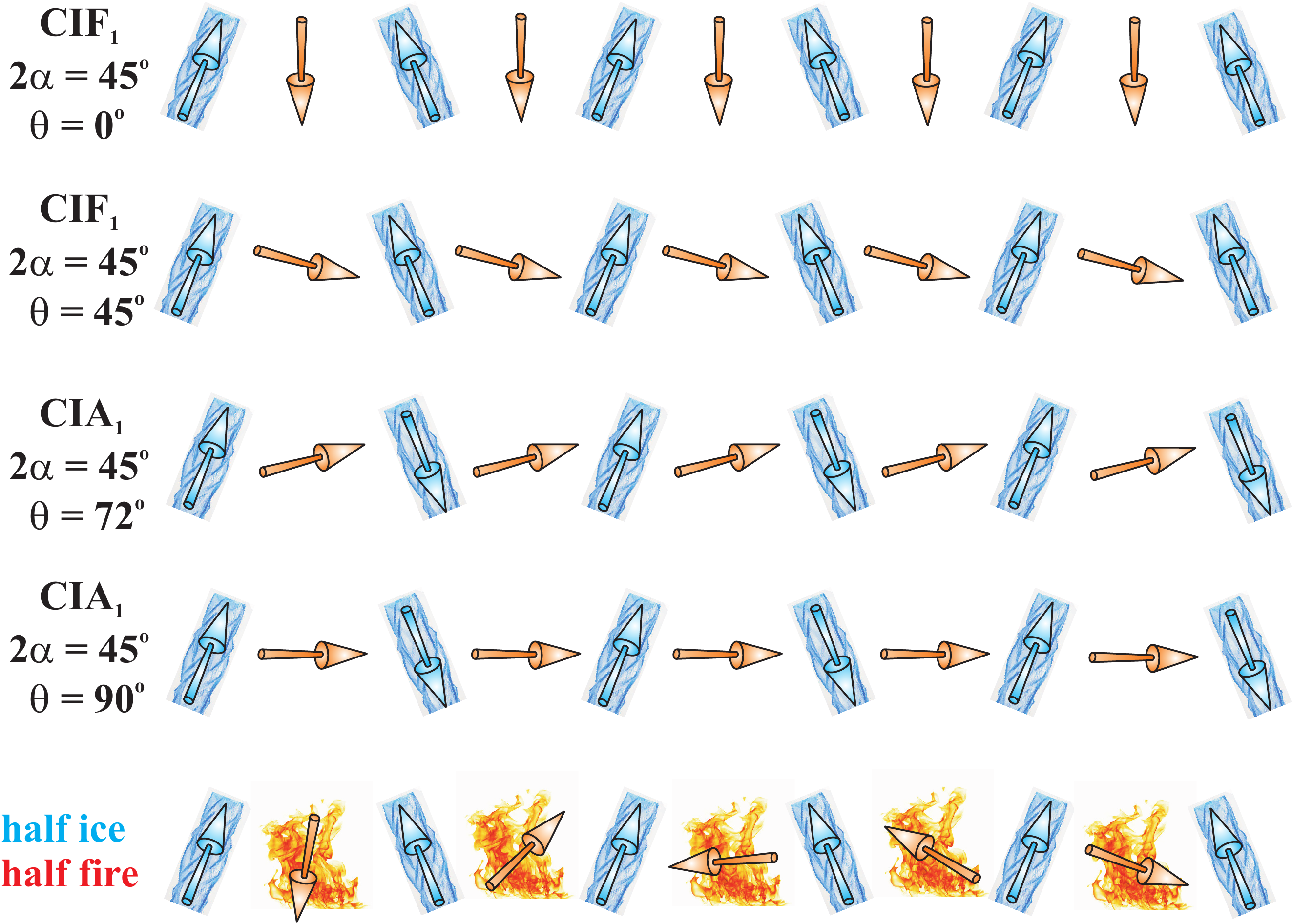}
\vspace{-0.4cm}
\caption{\label{con} (Color online) A schematic representation of mean values of the magnetic moments of the spin-1/2 Ising-Heisenberg chain for the fixed value of the canting angle $2\alpha = 45^{\circ}$, g-factors $g_{1}^{z_{1}}=g_{1}^{z_{2}}=20$, $g_{2}^{z}=g_{2}^{x}=2$ and selected spatial orientations of the external magnetic field $\theta = 0^{\circ}$, $45^{\circ}$, $72^{\circ}$ and $90^{\circ}$. The first two configurations correspond to the CIF$_1$ phase, the other two configurations to the CIA$_1$ phase and the last configuration correspond to a highly degenerate point called as 'half-ice, half-fire' ground state, where the Ising spins are fully frozen and the Heisenberg spins are completely randomly oriented.}
\end{figure}

First, we will comprehensively analyze all available ground states, which can be found by making use of the unit-cell eigenstates (\ref{eq:6}) and (\ref{eq:8}) with the eigenenergies (\ref{eq:5}) and (\ref{eq:7}) by considering all possible configurations of the Ising spins involved therein. In total, we can identify four different ground states, which will be classified according to a relative orientation of the Ising spins. More specifically, one finds two ground states CIF$_1$ and CIF$_2$ with the canted ferromagnetic alignment of the Ising spins
\begin{align}
\left|\mathrm{CIF}_{1}\right\rangle= & \prod_{i=1}^{N/2}\left|\psi^{-}\right\rangle_{2i-1}|\nearrow\rangle_{2i-1}\left
|\psi^{-}\right\rangle_{2i}\left|\nwarrow\right\rangle_{2i},\label{eq:39}\\
\left|\mathrm{CIF}_{2}\right\rangle= & \prod_{i=1}^{N/2}\left|\psi^{-}\right\rangle_{2i-1}|\swarrow\rangle_{2i-1}
\left|\psi^{-}\right\rangle_{2i}\left|\searrow\right\rangle_{2i},\label{eq:40}
\end{align}
and two ground states CIA$_1$ and CIA$_2$ with the canted antiferromagnetic alignment of the Ising spins
\begin{align}
\left|\mathrm{CIA}_{1}\right\rangle= & \prod_{i=1}^{N/2}\left|\psi^{-}\right\rangle_{2i-1}|\nearrow\rangle_{2i-1}
\left|\psi^{-}\right\rangle_{2i}\left|\searrow\right\rangle_{2i},\label{eq:41}\\
\left|\mathrm{CIA}_{2}\right\rangle= & \prod_{i=1}^{N/2}\left|\psi^{-}\right\rangle_{2i-1}|\swarrow\rangle_{2i-1}
\left|\psi^{-}\right\rangle_{2i}\left|\nwarrow\right\rangle_{2i}.\label{eq:42}
\end{align}
The state vector $\left|\nearrow\right>_{2i-1}$ ($\left|\swarrow\right>_{2i-1}$) corresponds to the spin state $\sigma_{2i-1}^{z_1} = 1/2$ ($\sigma_{2i-1}^{z_1} = -1/2$) of the odd-site Ising spin, the state vector $\left|\nwarrow\right>_{2i}$ ($\left|\searrow\right>_{2i}$) corresponds to the spin state $\sigma_{2i}^{z_2} = 1/2$ ($\sigma_{2i}^{z_2} = -1/2$) of the even-site Ising spin, whereas the odd- and even-site Heisenberg spins underlie a quantum superposition of both spin states according to Eqs. (\ref{eq:6}) and (\ref{eq:8}), respectively. It is quite obvious from Eqs. (\ref{eq:39}) and (\ref{eq:40}) that the CIF$_{2}$ phase is a mirror image of the CIF$_{1}$ phase with all over-turned spins and similarly, the CIA$_{2}$ phase is according to Eqs. (\ref{eq:41}) and (\ref{eq:42}) a mirror image of the CIA$_{1}$ phase. The former inter-relation between the CIF$_{1}$ and CIF$_{2}$ phases is a direct consequence of the symmetry of the model Hamiltonian (\ref{eq:1}) with respect to the magnetic field applied along $\theta = 90^{\circ}$ and $270^{\circ}$ directions, while the latter inter-relation between the CIA$_{1}$ and CIA$_{2}$ phases is a direct consequence of the symmetry of the model Hamiltonian (\ref{eq:1}) with respect to the magnetic field applied along $\theta = 0^{\circ}$ and $180^{\circ}$ directions. For illustration, a schematic representation of the mean values of the magnetic moments within the CIF$_1$ and CIA$_1$ ground states of the spin-1/2 Ising-Heisenberg chain is presented in Fig. \ref{con} for one particular value of the canting angle $2\alpha = 45^{\circ}$ between two anisotropy axes of the Ising spins.

\begin{figure}
\includegraphics[scale=0.21]{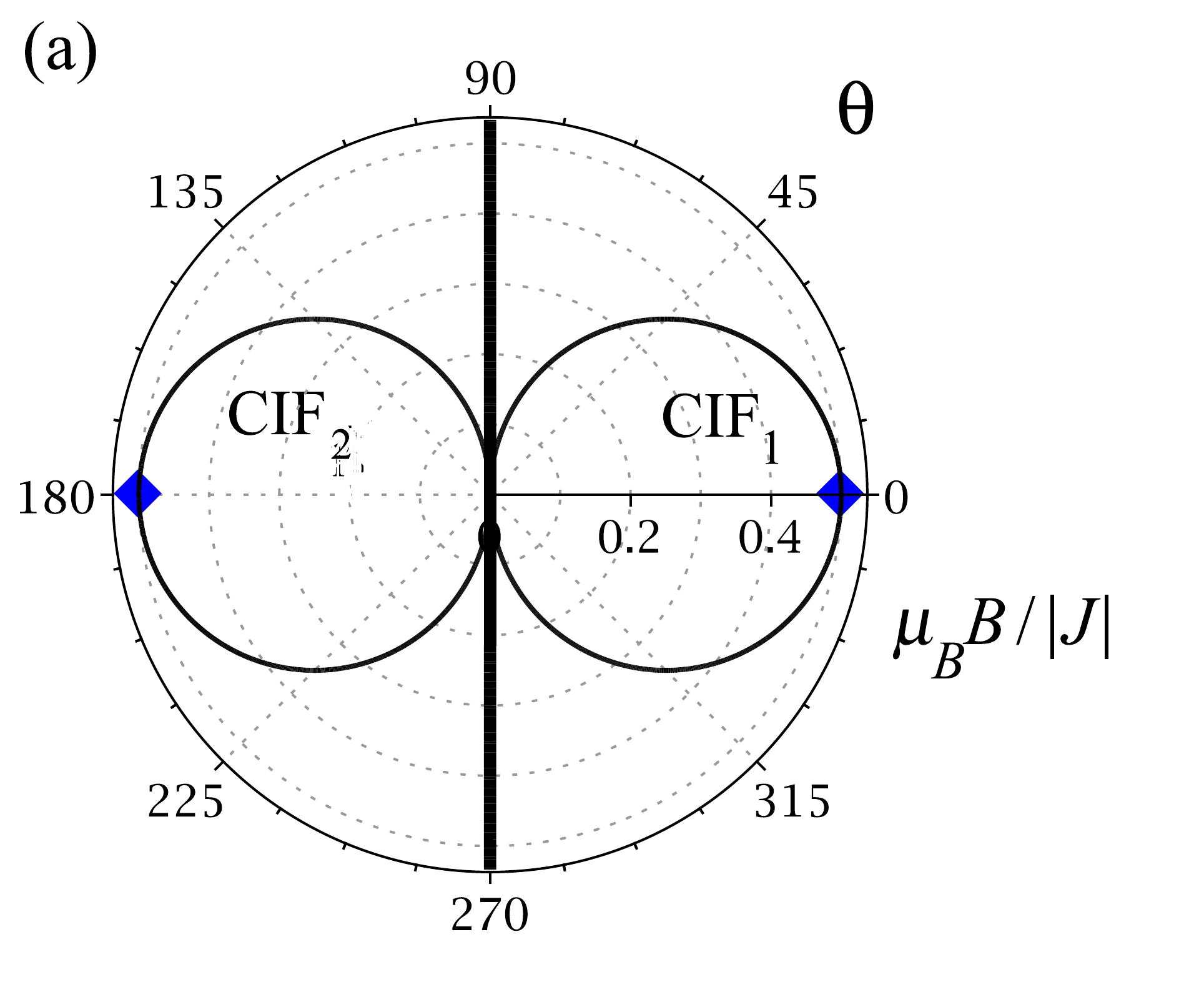}\includegraphics[scale=0.21]{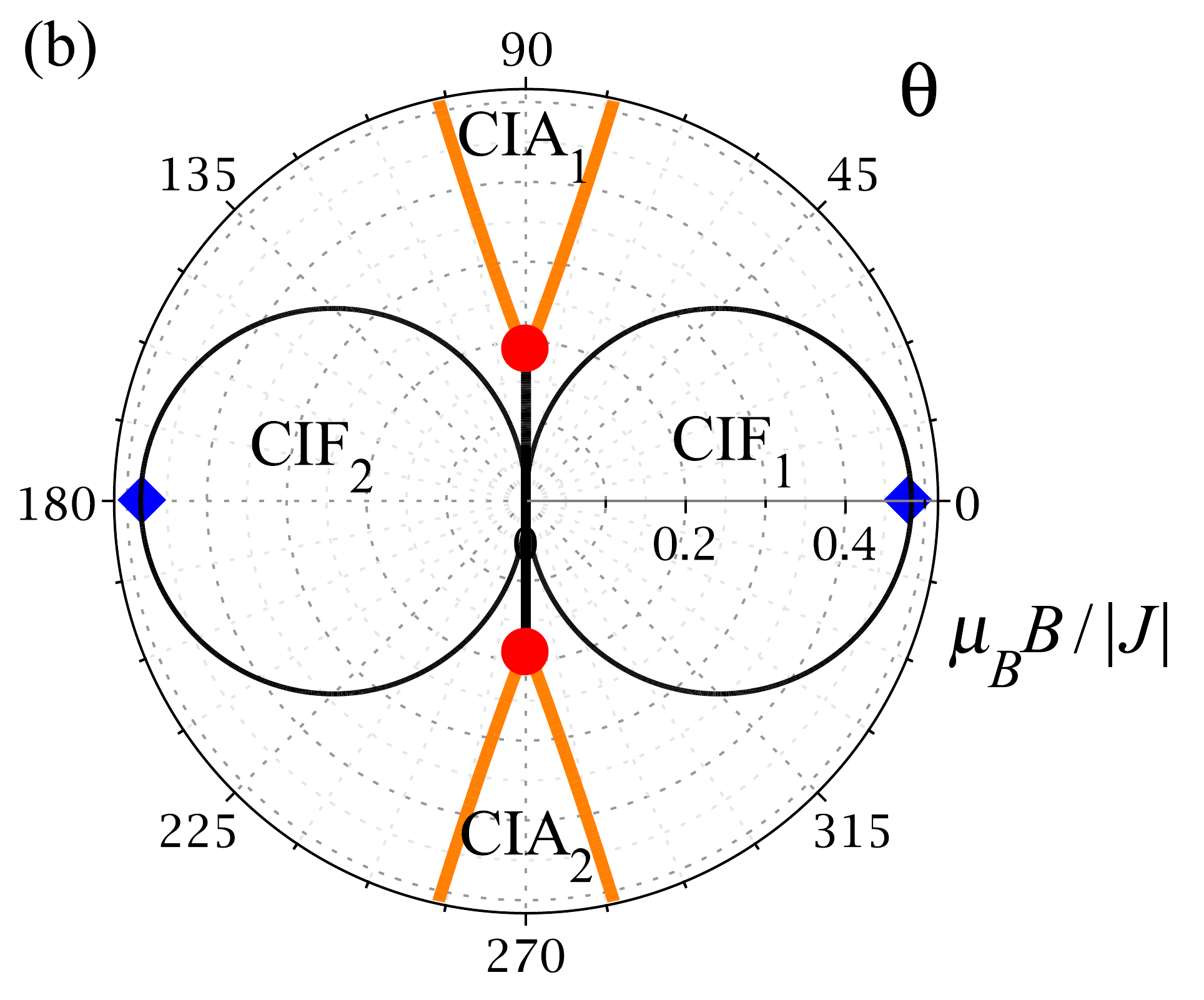}
\includegraphics[scale=0.21]{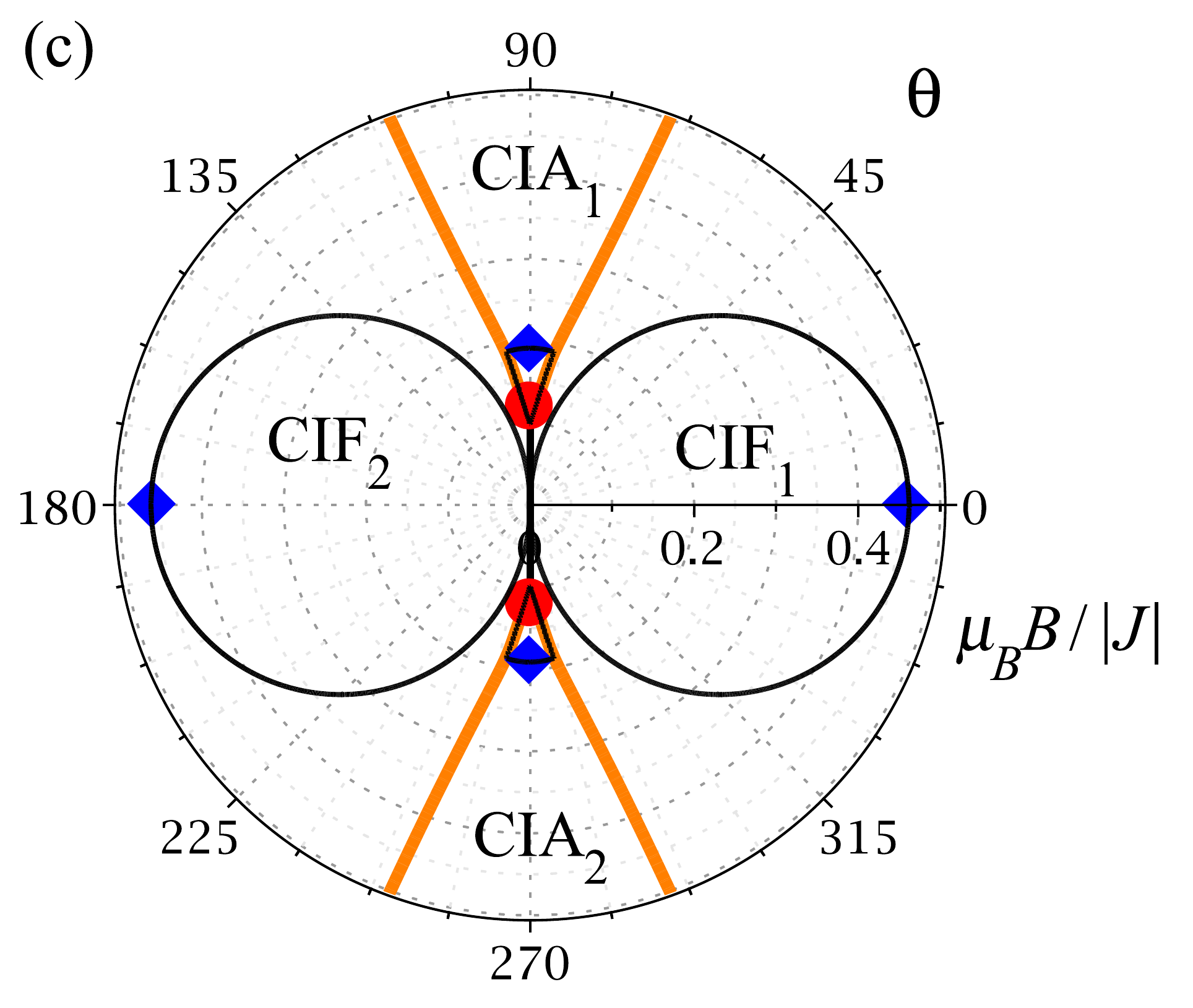}\includegraphics[scale=0.21]{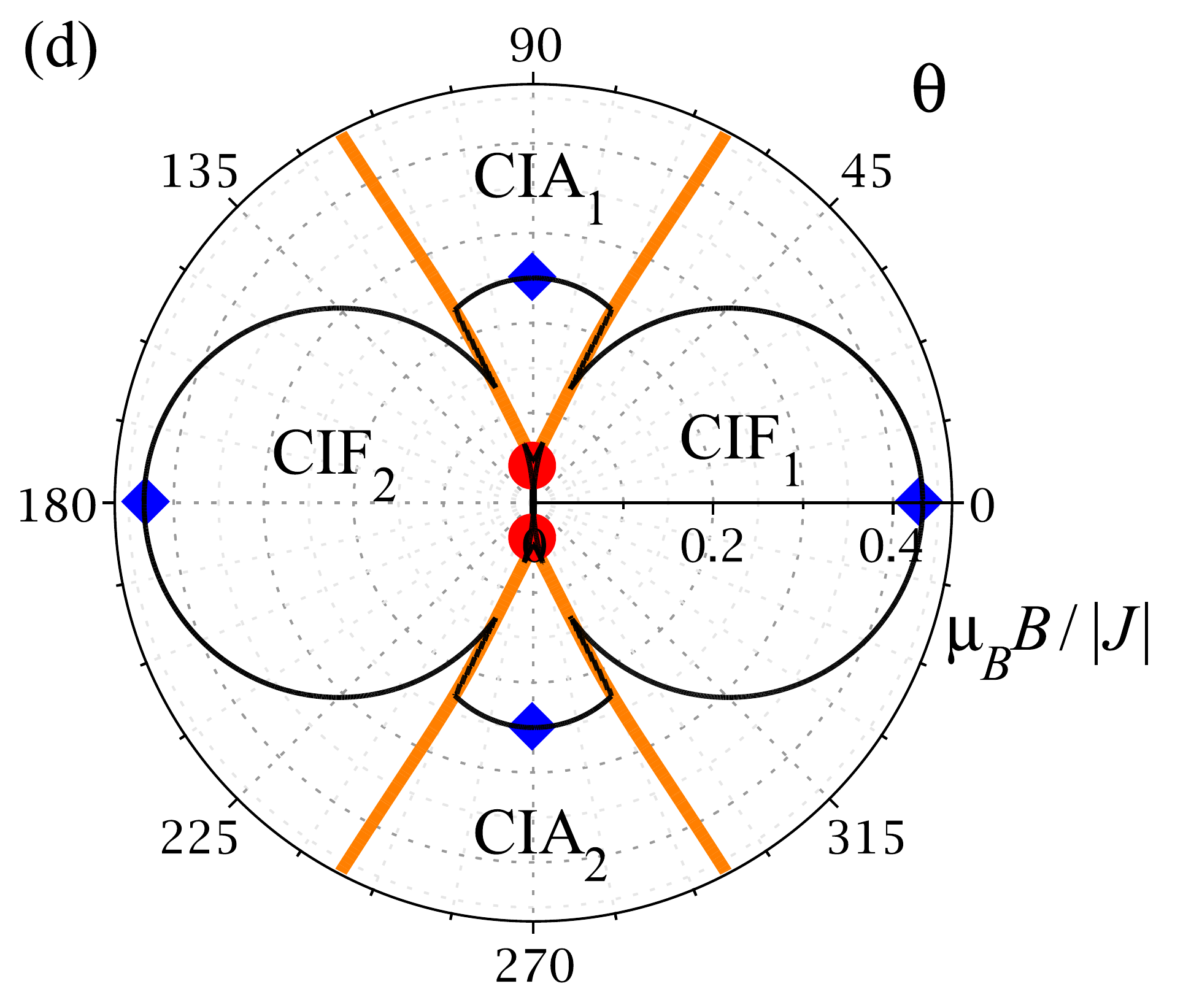}
\includegraphics[scale=0.21]{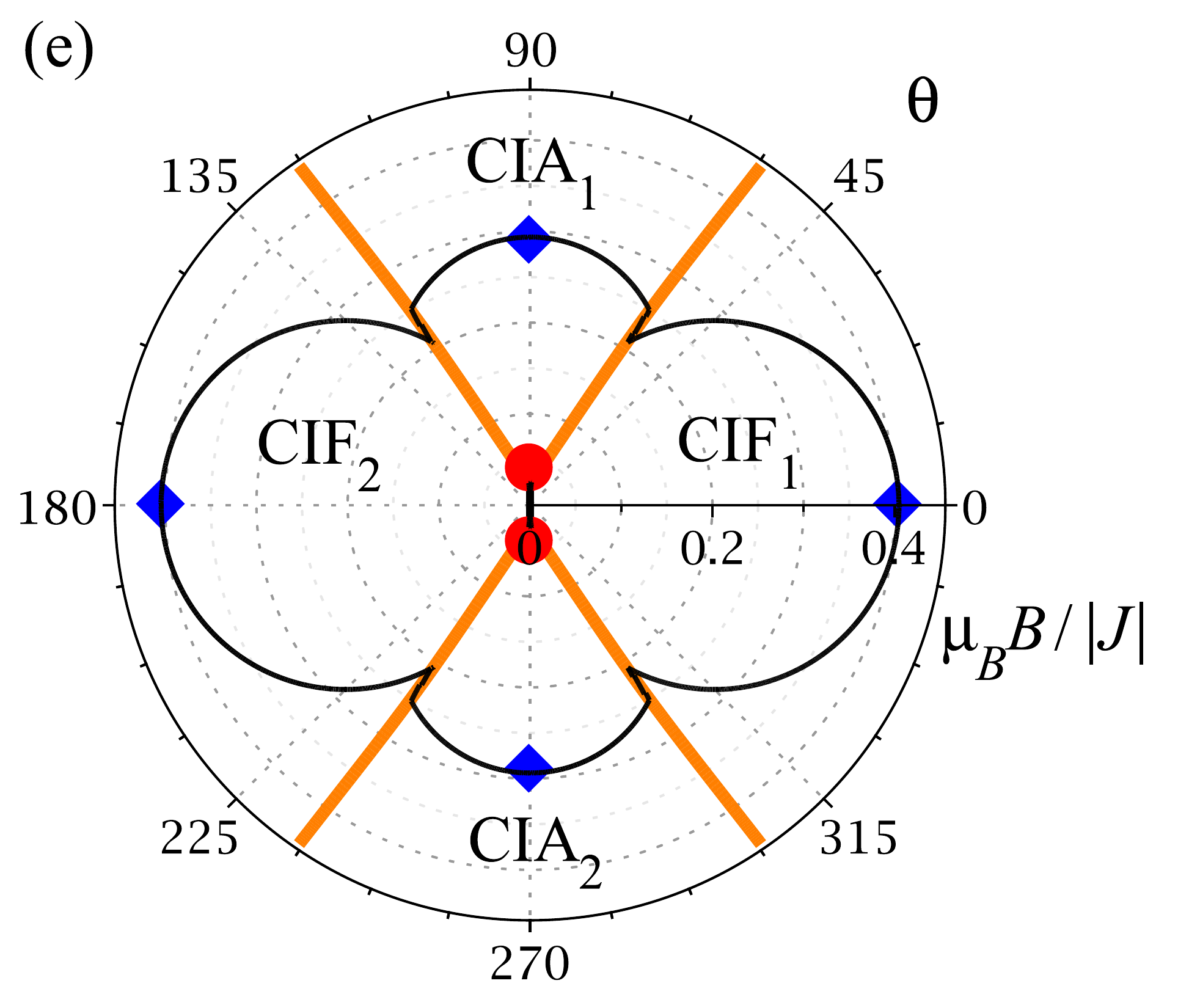}\includegraphics[scale=0.21]{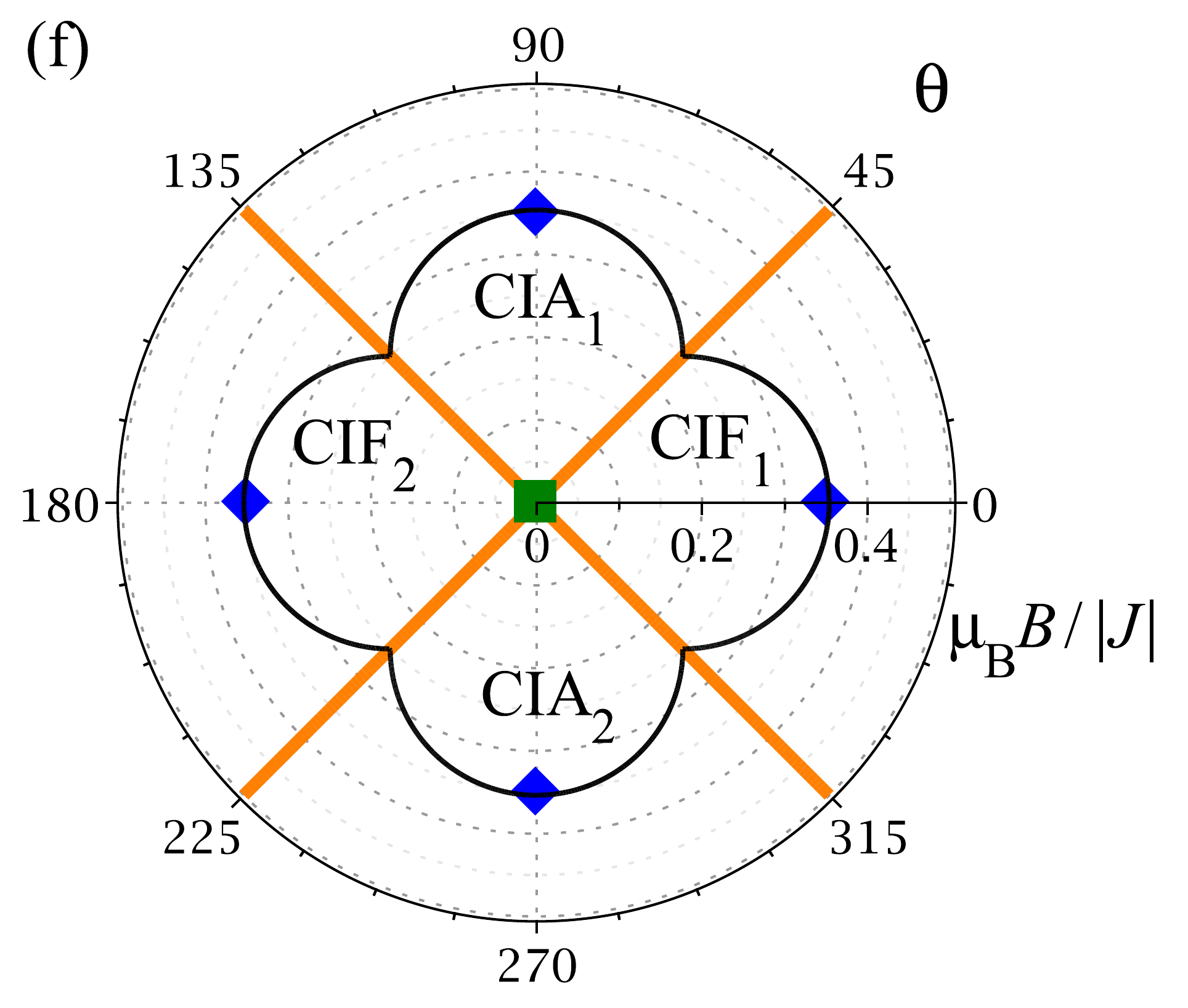}
\vspace{-0.7cm}
\caption{\label{fig:2} (Color online) The ground-state phase diagrams in polar coordinate system, where the radius represents a relative size of the magnetic field $\mu_{\rm B}B/|J|$ and the polar angle $\theta$ determines an orientation of magnetic field with respect to the global frame $z$-axis for the fixed values of g-factors $g_{1}^{z_{1}}=g_{1}^{z_{2}}=20$, $g_{2}^{z}=g_{2}^{x}=2$ and several values of the canting angle $2\alpha$: (a) $2\alpha=0^{\circ}$, (b) $2\alpha=30^{\circ}$, (c) $2\alpha=45^{\circ}$, (d) $2\alpha=60^{\circ}$, (e) $2\alpha=72^{\circ}$, (f) $2\alpha=90^{\circ}$.}
\end{figure}

A few typical ground-state phase diagrams are plotted in Fig. \ref{fig:2} in a polar coordinate system ($\mu_{\rm B}B/|J|$, $\theta$) for several values of the canting angle $2\alpha$ between two different co-planar anisotropy axes. It is noteworthy that the radius of polar coordinates determines a relative size of the external magnetic field $\mu_{\rm B} B/|J|$ and the angular coordinate is identified with the azimuthal angle $\theta$ determining inclination of the external magnetic field with respect to the global frame $z$-axis.

It is clear from Fig. \ref{fig:2} that all displayed ground-state phase diagrams actually have an evident symmetry with respect to $\theta = 0^{\circ}$, $90^{\circ}$, $180^{\circ}$ and $270^{\circ}$ axes, so our further discussion may be restricted without loss of generality only to a first quadrant $\theta \in [0^{\circ}, 90^{\circ}]$. It can be found that CIF$_1$ and CIF$_2$ phases then coexist together along a macroscopically non-degenerate phase boundary (thick black line) lying at $\theta=90^{\circ}$, which terminates at a triple coexistence point (red circle) of CIF$_1$, CIF$_2$ and CIA$_1$ phases with a residual entropy $S=N k_{\rm B} \ln[(\sqrt{5}+3)/2]/2$ (see Appendix A for details on the macroscopic degeneracy at this special point). On the other hand, the coexistence (thick orange) line between CIF$_{1}$ and CIA$_{1}$ phases is macroscopically degenerate with the residual entropy $S= N k_{\rm B} \ln(2)/2$ due to a paramagnetic character of a half of the Ising spins. The Heisenberg spins are completely free to flip at macroscopically degenerate points (blue diamonds) with the finite residual entropy $S= N k_{\rm B} \ln(2)$ given by the coordinates $\mu_{\rm B}B/|J| = \cos(\alpha)/2$, $\theta=0^{\circ}$ and $\mu_{\rm B}B/|J|=\sin(\alpha)/2$, $\theta=90^{\circ}$ for $\alpha \gtrsim 20^{\circ}$. The macroscopic degeneracy at these special points correspond to a novel type of spin frustration called as 'half ice, half fire' \cite{yin15,tor17,oha17}, which originates from a difference between g-factors being responsible for a fully frozen (ordered) character of the Ising spins and a fully random (disordered) character of the Heisenberg spins. Finally, a thin black curve determines a contour plot for a zero projection of the magnetization of the Heisenberg spins $m_2^x\sin(\theta)+m_2^z\cos(\theta)=0$ towards the external magnetic field, along which their contribution to the total magnetization vanishes. It is worthwhile to remark that the contribution of the Heisenberg spins diminishes (reinforces) the total magnetization inside (outside) of a parameter region delimited by this contour line.

Fig.~\ref{fig:2}(a) depicts the ground-state phase diagram for the very special case $2\alpha=0^{\circ}$ without canting of the local anisotropy axes. It is quite obvious that only the phases CIF$_1$ and CIF$_2$ with a uniform orientation of the Ising spins are present in the ground-state phase diagram of this particular case. The blue diamond located at $[\mu_BB/|J|, \theta]$=$[0.5, 0^{\circ}]$ denotes a macroscopically degenerate point, which corresponds to an abrupt magnetization jump in a zero-temperature magnetization curve closely connected to a field-induced flip of the Heisenberg spins. On the other hand, the canting of the local anisotropy axes of the Ising spins is responsible for presence of two striking phases CIA$_1$ and CIA$_2$ with a staggered character of the Ising spins [see Fig.~\ref{fig:2}(b)-(f)]. As one can see, the phases CIA$_1$ and CIA$_2$ spread over a wider parameter space at the expense of the phases CIF$_1$ and CIF$_2$ as the canting angle $2 \alpha$ increases. If the external magnetic field is applied along the $z$-axis ($\theta = 0^{\circ}$) lying in the middle of two spatial directions determining the anisotropy axes of the Ising spins, then, the field-induced flip of the Heisenberg spins shifts towards lower magnetic fields upon increasing the canting angle $2 \alpha$ [see Fig.~\ref{fig:2}(b)-(f)]. The field-induced flip of the Heisenberg spins may also take place when the external magnetic field is applied perpendicular to the $z$-axis (i.e. $\theta = 90^{\circ}$) as long as the canting angle $2\alpha$ is sufficiently high. However, the relative size of the magnetic field required for a relevant spin flip along the spatial direction $\theta = 90^{\circ}$ contrarily rises upon increasing the canting angle $2\alpha$. Finally, it is noteworthy that a sharp field-induced flip of the Heisenberg spins generally becomes smoother when the external magnetic field is applied along other spatial directions. Under this condition, the field-driven reorientation of the Heisenberg spins can be found along a thin black contour line, where the contribution of the Heisenberg spins to the total magnetization vanishes.

\subsection{Magnetization process}
Let us begin with a detailed analysis of the zero-temperature magnetization process of the spin-1/2 Ising-Heisenberg chain in an arbitrarily oriented magnetic field. For this purpose, one has to determine first the ground-state energy $E_{0}=\min(E_{CIF_{1}},E_{CIF_{2}},E_{CIA_{1}},E_{CIA_{2}})$ as a minimum eigenenergy out of four available lowest-energy eigenstates given by Eqs. (\ref{eq:39})-(\ref{eq:42}). Then, two orthogonal projections of the magnetization of the Heisenberg spins can be straightforwardly calculated according to
\begin{align}
\begin{alignedat}{1}m_{2}^{x}=-g_2^x \mu_{\rm B} \frac{\partial E_{0}}{\partial h^{x}}\end{alignedat}
\quad\text{and}\quad & m_{2}^{z}=-g_2^z \mu_{\rm B}\frac{\partial E_{0}}{\partial h^{z}},
\label{eq:48}
\end{align}
while the magnetizations of two inequivalent Ising spins along their individual quantization axes follow from
\begin{align}
\begin{alignedat}{1}m_{1}^{z_1}=-g_1^{z_1} \mu_{\rm B}\frac{\partial E_{0}}{\partial h^{z_1}}\end{alignedat}
\quad\text{and}\quad & m_{1}^{z_2}=-g_1^{z_2} \mu_{\rm B}\frac{\partial E_{0}}{\partial h^{z_2}}.
\label{eq:49}
\end{align}
\begin{figure*}
\includegraphics[scale=0.5]{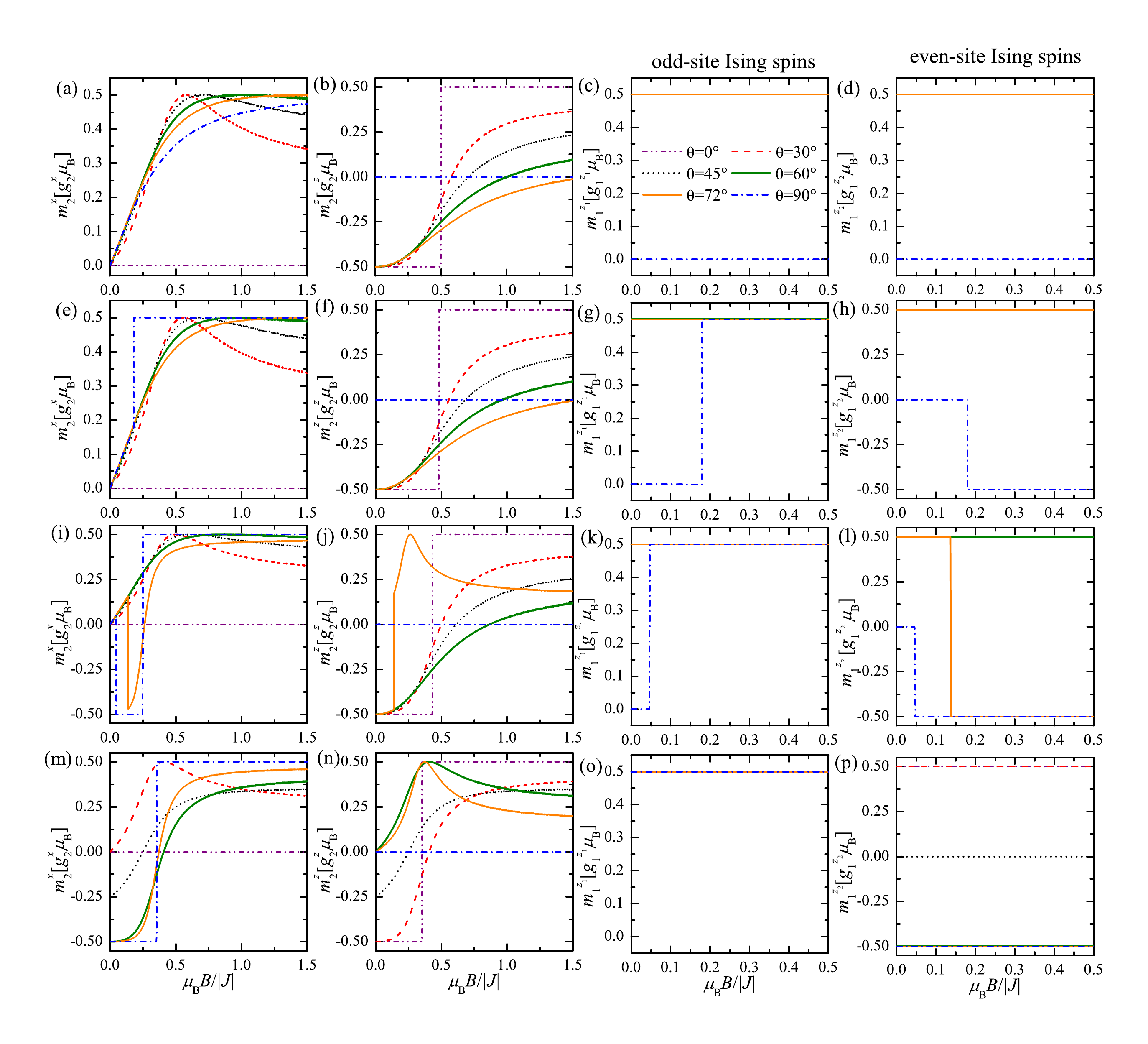}
\caption{\label{fig:3} (Color online) Two orthogonal projections of the zero-temperature magnetization of the Heisenberg spins ($m_{2}^{x}$, $m_{2}^{z}$) and the zero-temperature magnetizations of two topologically different Ising spins ($m_{1}^{z_{1}}$, $m_{1}^{z_{2}}$) as a function of the dimensionless magnetic field $\mu_{\rm B} B/|J|$ for the fixed values of g-factors $g_{1}^{z_{1}}=g_{1}^{z_{2}}=20$, $g_{2}^{x}=g_{2}^{z}=2$ and four different values of the canting angle: (a)-(d) $2\alpha=0^{\circ}$; (e)-(h) $2\alpha=30^{\circ}$; (i)-(l) $2\alpha=60^{\circ}$; (m)-(p) $2\alpha=90^{\circ}$. }
\end{figure*}
The total magnetization $m_t$ can be calculated according to Eq. (\ref{eq:31}) from the aforementioned four components, which are plotted in Fig.~\ref{fig:3} as a function of the relative strength of the magnetic field $\mu_{\rm B} B/|J|$ for several values of the canting angle $2\alpha$. The first (second) column shows a transverse (longitudinal) component of the magnetization of the Heisenberg spins $m_{2}^{x}$ ($m_{2}^{z}$), while the third (fourth) column illustrates the magnetization of the odd(even)-site Ising spins $m_{1}^{z_{1}}$ ($m_{1}^{z_{2}}$) along their local anisotropy axis $z_{1}$ ($z_{2}$). It is worth noticing that the overall magnetization of the Heisenberg spins acquires its maximum value $\sqrt{(m_{2}^{x})^{2}+(m_{2}^{z})^{2}} = g_2 \mu_{\rm B}/2$ ($g_2 = g_{2}^{x} = g_{2}^{z}$) for all parameters except those driving the investigated system at phase boundaries between different ground states.

The plots shown in Fig.~\ref{fig:3}(a-d) illustrate the magnetization scenario for the specific case with zero canting angle $2\alpha=0^{\circ}$. As far as the magnetization of the Heisenberg spins is concerned, the transverse component $m_{2}^{x}$ becomes zero and the longitudinal component $m_{2}^{z}$ fully saturated whenever the magnetic field is oriented along a unique anisotropy axis of the Ising spins [see dash-dot-dot line in Fig.~\ref{fig:3}(a-b) for $\theta=0^{\circ}$]. The magnetization of the Ising spins is at the same time fully saturated, whereas the magnetizations of the Ising and Heisenberg spins are oriented in opposite (parallel) to each other at sufficiently low (high) magnetic fields. If the magnetic field is misaligned from a unique anisotropy axis of the Ising spins $\theta \neq 0^{\circ}$, then, the transverse magnetization of the Heisenberg spins displays a non-monotonous field dependence with a single global maximum at a moderate magnetic field that coincides with zero value of the longitudinal magnetization of the Heisenberg spins. The longitudinal magnetizations of the Ising and Heisenberg spins become identically equal to zero for the the external magnetic field oriented along the $x$-axis (i.e. $\theta=90^{\circ}$), whereas the transverse magnetization of the Heisenberg spins then rises steadily with increasing of the magnetic field.

The magnetization scenario for relatively small canting angles between both local anisotropy axes is illustrated in Fig.~\ref{fig:3}(e-h) on a particular example with $2\alpha=30^{\circ}$. It can be observed from Fig.~\ref{fig:3}(f) that the longitudinal magnetization of the Heisenberg spins shows qualitatively the same behavior as described previously for the case $2\alpha=0^{\circ}$ irrespective of the field orientation $\theta$. The transverse magnetization of the Heisenberg spins exhibits qualitatively different behavior just for the spatial orientation $\theta=90^{\circ}$, under which an abrupt magnetization jump occurs instead of a smooth field dependence. Interestingly, the magnetizations of odd- and even-site Ising spins display at the same magnetic field an abrupt jump from zero magnetization, the former magnetization shows a discontinuous jump to the highest and the latter one to the lowest possible value. Apparently, the observed magnetization jump can be attributed to a discontinuous field-driven phase transition from the CIF$_1$ phase towards the  CIA$_1$ phase with a staggered character of the Ising spins.

The magnetic-field dependencies of individual sublattice magnetizations are depicted in Fig.~\ref{fig:3}(i-l) for the moderate value of the canting angle $2\alpha=60^{\circ}$. As far as two components of the magnetization of the Heisenberg spins are concerned, the notable differences in a profile of zero-temperature magnetization curves are evident just for large tilting angles of the external magnetic field $\theta=72^{\circ}$ and $90^{\circ}$ though the observed magnetization jumps can be again related to the discontinuous field-induced phase transition between the CIF$_1$ and CIA$_1$ phases. It should be pointed out, however, that shortly after the relevant field-driven phase transition the longitudinal (transverse) component of the magnetization of the Heisenberg spins reaches for $\theta=72^{\circ}$ a local maximum (minimum), which is successively followed by a gradual decline (rise) upon strengthening of the external magnetic field.

Last but not least, the magnetization scenario for the large canting angle $2\alpha=90^{\circ}$ between both local anisotropy axes is shown in Fig.~\ref{fig:3}(m-p). While the magnetization of odd-site Ising spins always acquires its maximum value, the magnetization of even-site Ising spins acquires either its highest or lowest possible value depending on whether the tilting angle of external magnetic field is smaller or greater than $\theta=45^{\circ}$. This result would suggest that the CIA$_1$ phase persists down to very low fields whenever the magnetic field is applied at large tilting angles $\theta>45^{\circ}$. Another interesting observation is that the longitudinal and transverse components of the magnetization of the Heisenberg spins are complementary to each other. As a matter of fact, the longitudinal magnetization of the Heisenberg spins at a given tilting angle $\theta$ is identical with the transverse magnetization of the Heisenberg spins at a complementary tilting angle $90^{\circ} - \theta$. Therefore, both components of the magnetization of the Heisenberg spins become equal for $\theta=45^{\circ}$ due to the symmetry reason, whereas the magnetization of even-site Ising spins then exceptionally equals zero. These results are in concordance with a mutual coexistence of the CIF$_1$ and CIA$_1$ phases along their phase boundary at $\theta=45^{\circ}$.

With this background, let us discuss zero-temperature variations of the total magnetization $m_{t}$ with the external magnetic field. The angular dependence of the total magnetization $m_t$ of the spin-1/2 Ising-Heisenberg chain can be obtained either from the aforedescribed magnetization components according to Eq. (\ref{eq:31}) or as the field derivative of the ground-state energy
\begin{equation}
m_{t}=-\frac{\partial E_{0}}{\partial B}.\label{eq:50}
\end{equation}
\begin{figure}
\includegraphics[scale=0.28]{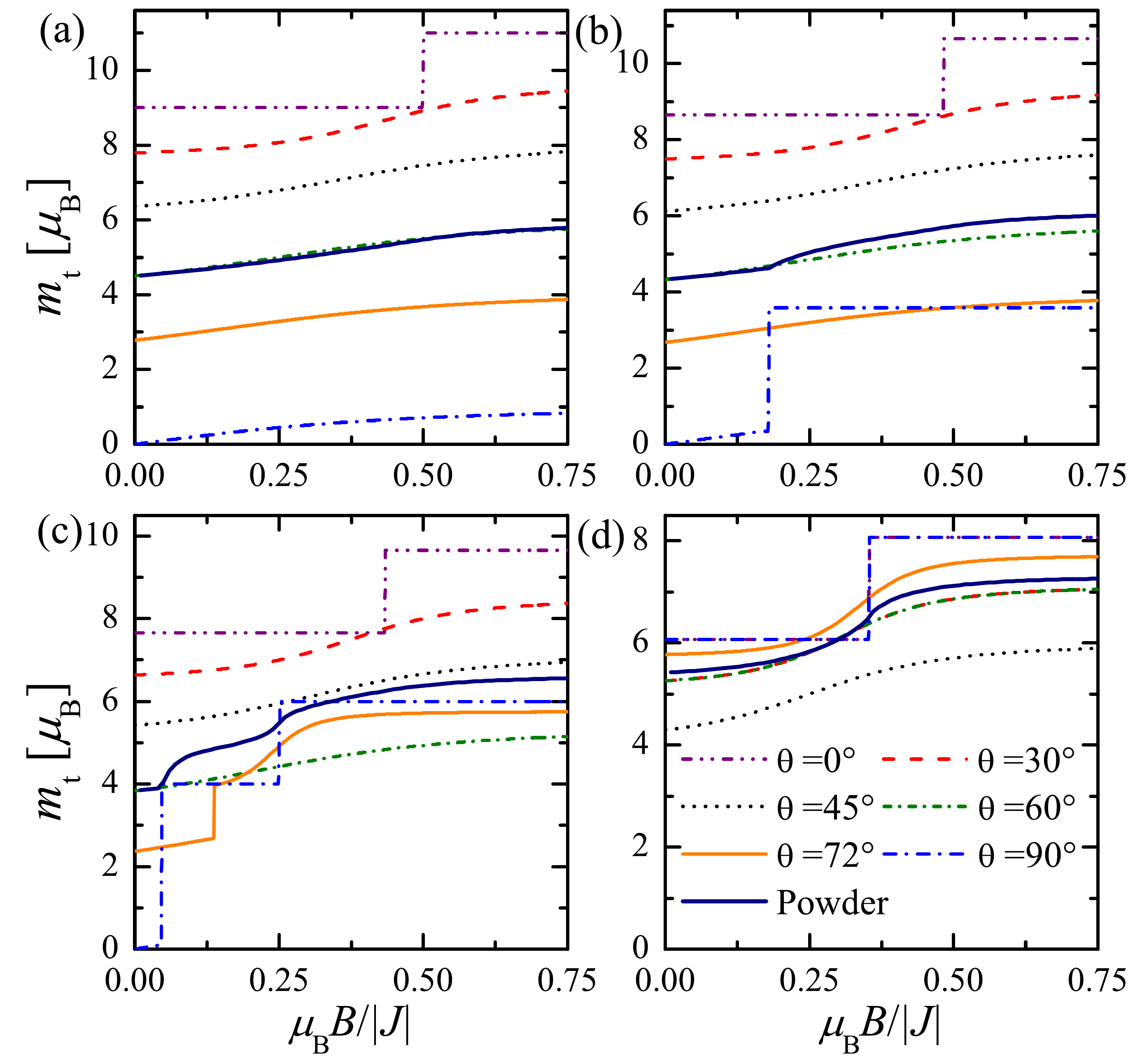}
\caption{\label{fig:4} (Color online) The total magnetization $m_{t}$ against the dimensionless magnetic field $\mu_{\rm B} B/|J|$ at zero temperature for the fixed values of g-factors $g_{2}^{z}=g_{2}^{x}=2$, $g_{1}^{z_{1}}=g_{1}^{z_{2}}=20$, a few different spatial orientations $\theta$ of the magnetic field and four different canting angles: (a) $2\alpha=0^{\circ}$, (b) $2\alpha=30^{\circ}$, (c) $2\alpha=60^{\circ}$ and (d) $2\alpha=90^{\circ}$. The powder magnetization $m_{powder}$ calculated according to Eq. (\ref{eq:31}) is also plotted for the sake of comparison. The same line code is used in all plots.}
\end{figure}
The total magnetization is plotted in Fig.~\ref{fig:4} as a function of the magnetic field $\mu_{\rm B} B/|J|$ for a few different spatial orientations of the magnetic field ($\theta$) and four different canting angles $2 \alpha$. It is evident that the zero-temperature magnetization curves display true magnetization plateaus for the magnetic field oriented along the $z$-axis (i.e. $\theta = 0^{\circ}$) and the quasi-plateaus for relatively high tilting angles close to $\theta \approx 90^{\circ}$. The locus of all discontinuous magnetization jumps closely connected either with a breakdown of the magnetization plateaus or quasi-plateaus is fully consistent with presence of the field-driven phase transitions comprehensively  discussed at the ground-state analysis. It should be stressed, however, that the total magnetization exhibits for most of the field orientations a smooth dependence on a magnetic field, which relates to local quantum fluctuations introduced by the external magnetic field whenever it is misaligned from the anisotropy axes of the Ising spins. Note furthermore that the powder magnetization also exhibits at zero temperature a smooth field dependence, which originates from the angular averaging (\ref{eq:30}) embracing an integration over all possible field orientations.

\begin{figure}
\includegraphics[scale=0.24]{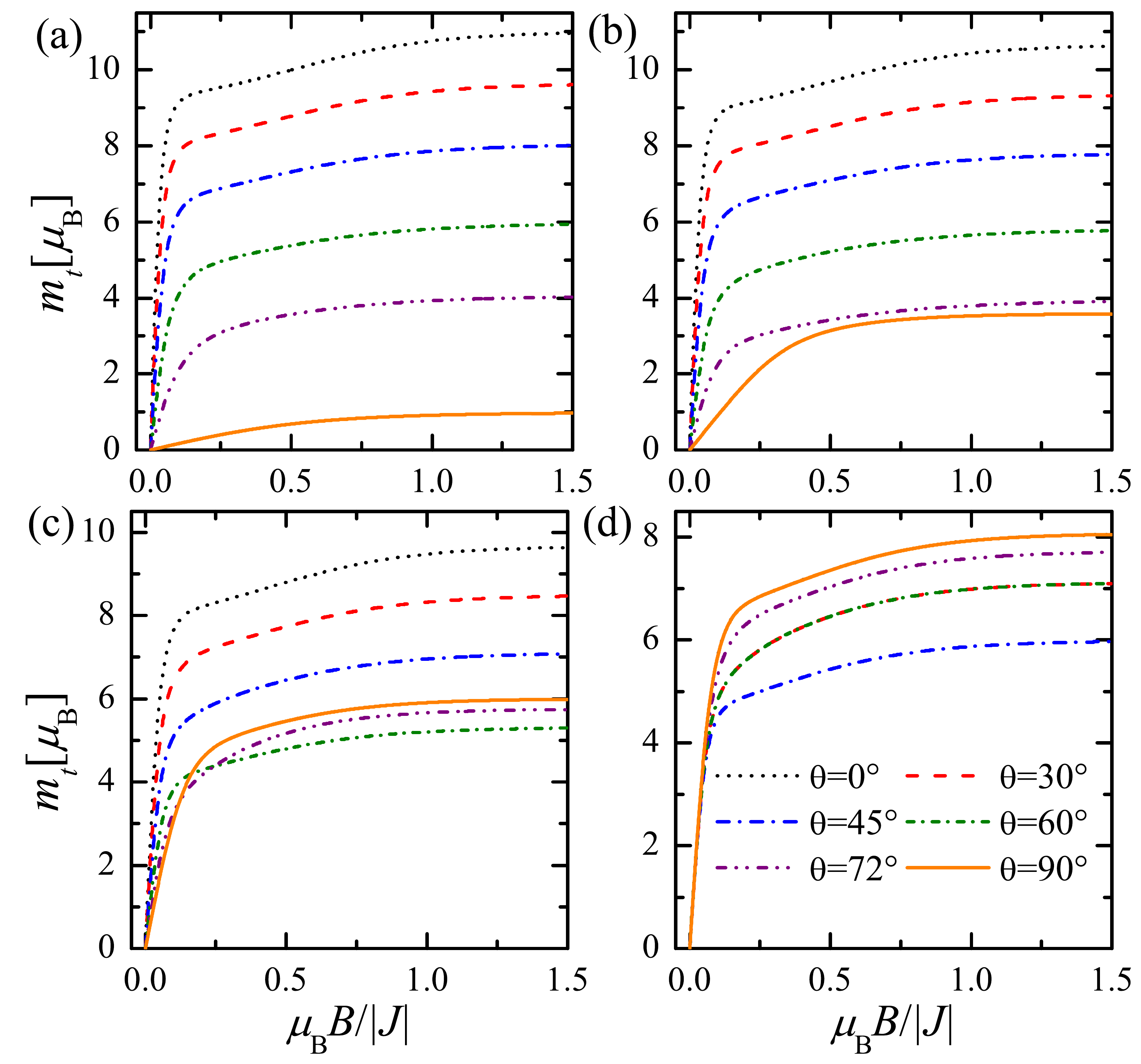}
\caption{\label{fig:5} (Color online) The total magnetization $m_{t}$ as a function of the dimensionless magnetic field $\mu_{\rm B} B/|J|$ at finite temperature $k_{\rm B} T/J = 0.5$ for the fixed values of g-factors $g_{2}^{z}=g_{2}^{x}=2$, $g_{1}^{z_{1}}=g_{1}^{z_{2}}=20$, a few different spatial orientations $\theta$ of the magnetic field and four different canting angles: (a) $2\alpha=0^{\circ}$, (b) $2\alpha=30^{\circ}$, (c) $2\alpha=60^{\circ}$ and (d) $2\alpha=90^{\circ}$. The powder magnetization $m_{powder}$ calculated according to Eq. (\ref{eq:31}) is also plotted for the sake of comparison. The same line code is used in all plots.}
\end{figure}

Next, the finite-temperature magnetization curves are depicted in Fig.~\ref{fig:5} for several spatial orientations of the magnetic field $\theta$ and a few different canting angles $2\alpha$ between the local anisotropy axes of the Ising spins with the aim to bring insight into a mutual interplay of quantum and thermal fluctuations. As one can see, the magnetization curves generally display at finite temperatures smoother field dependence, whereas the sharp magnetization jumps observable at zero temperature round off upon increasing temperature. In addition, it can be understood from Fig.~\ref{fig:5} that the saturation value of the total magnetization is progressively suppressed upon increasing the canting angle $2\alpha$ between the local anisotropy axes.

\begin{figure}
\includegraphics[scale=0.24]{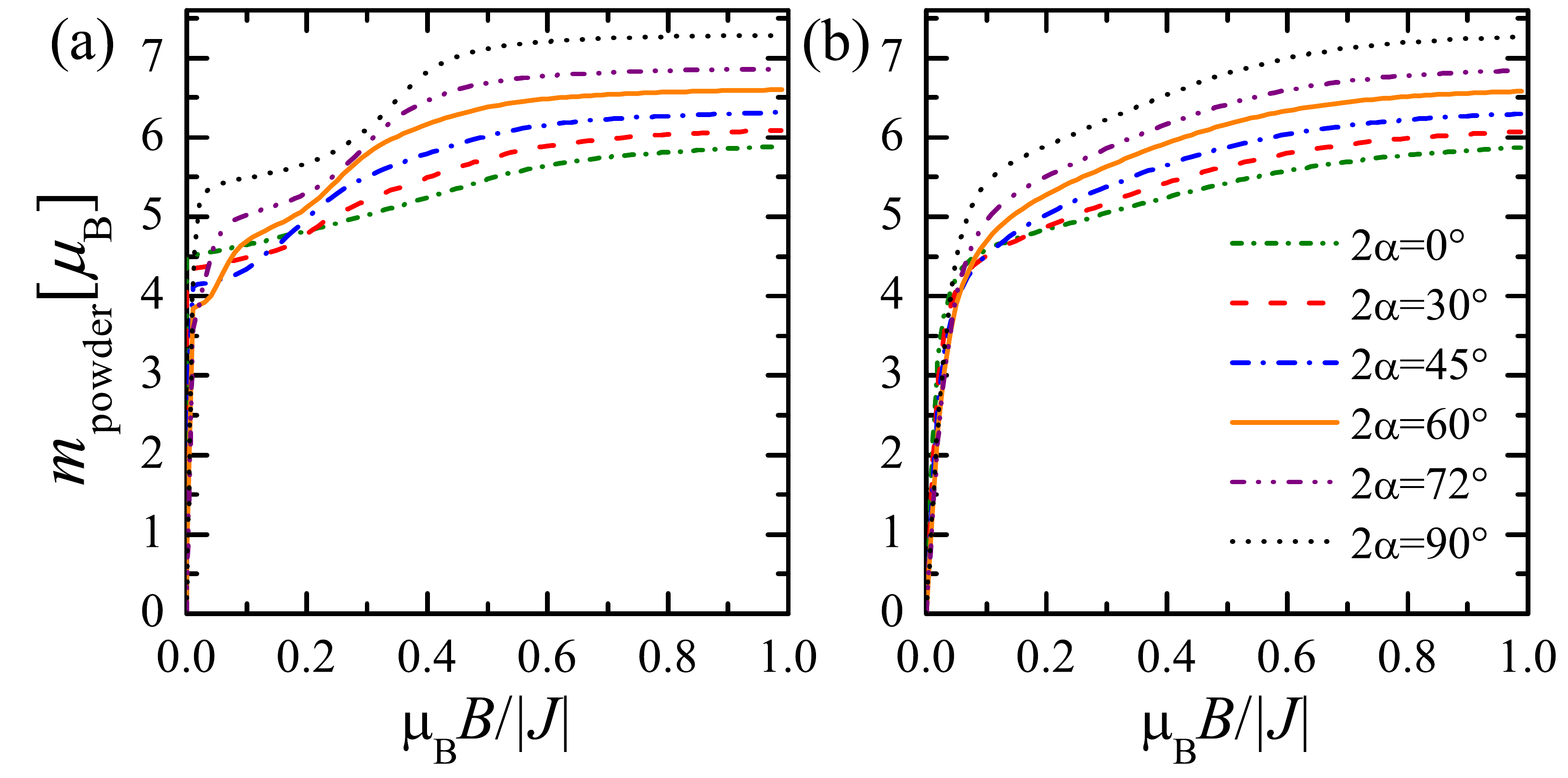}
\caption{\label{fig:6} (Color online) The powder magnetization as a function of the magnetic field $\mu_{\rm B}B/|J|$ for the fixed values of g-factors  $g_{1}^{z_{1}}=g_{1}^{z_{2}}=20$, $g_{2}^{z}=g_{2}^{x}=2$, several values of the canting angle $2\alpha$ between the anisotropy axes and two different temperatures: (a) $k_{\rm B}T/|J|=0.0$; (b) $k_{\rm B}T/|J|=0.5$.}
\end{figure}

The powder magnetization shown in Fig.~\ref{fig:6} captures magnetization curves of polycrystalline samples, which are subject to an integration over all possible orientations of the magnetic field with respect to the anisotropy axes due to a random orientation of individual crystallites. It is quite evident that the powder averaging has similar effect as the rising temperature, because the magnetization plateaus, quasi-plateaus and jumps are in general rounded as a consequence of an integration over distinct relative orientations of the external magnetic field with respect to the local anisotropy axes (see Figs.~\ref{fig:4} and \ref{fig:5} for comparison). Moreover, it is worthwhile to remark that the saturation value of the powder magnetization contraintuitively displays a small increment upon increasing the canting angle between the anisotropy axes of the Ising spins.

\subsection{Susceptibility}

\begin{figure*}[!t]
\includegraphics[scale=0.5]{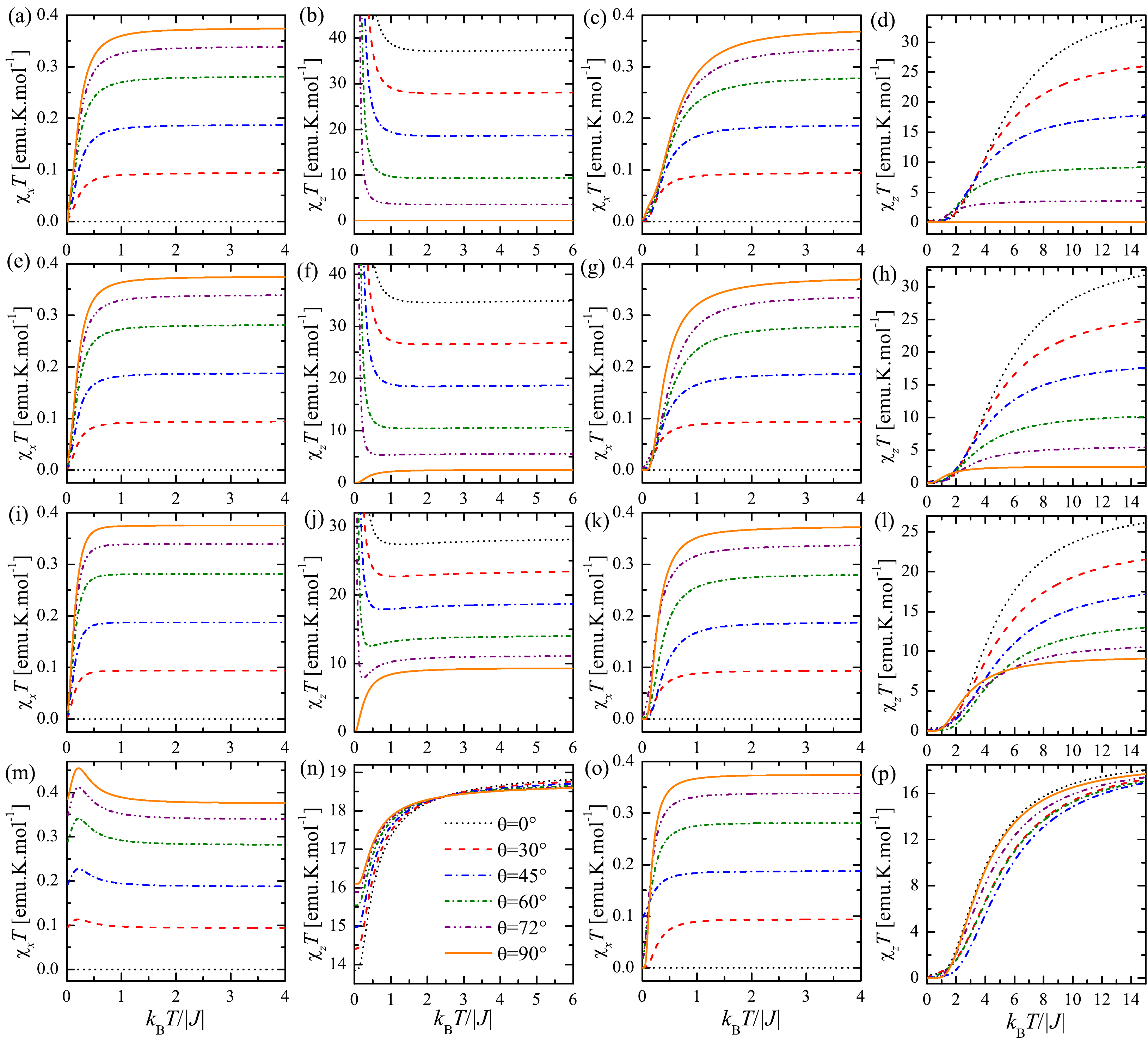}
\caption{\label{fig:8} (Color online) The transverse ($\chi_x T$) and longitudinal ($\chi_z T$) components of the susceptibility times temperature product versus dimensionless temperature $k_{\rm B}T/|J|$ for the fixed values of g-factors $g_{1}^{z_{1}}=g_{1}^{z_{2}}=20$, $g_{2}^{z}=g_{2}^{x}=2$, several values of the tilting angle $\theta$ and four different canting angles: (a)-(d) $2\alpha=0^{\circ}$; (e)-(h) $2\alpha=30^{\circ}$; (i)-(l) $2\alpha=60^{\circ}$; (m)-(p) $2\alpha=90^{\circ}$. The first two columns refer to the susceptibility times temperature product at zero magnetic field $\mu_{\rm B} B/|J|=0$, while another two columns refer to the susceptibility times temperature product at the finite magnetic field $\mu_{\rm B} B/|J|=0.5$.}
\end{figure*}

The transverse and longitudinal components of the susceptibility times temperature product are plotted in Fig.~\ref{fig:8} against temperature for several values of the tilting angle $\theta$ and four different canting angles $2\alpha$. The first two columns display thermal variations of the initial (zero-field) susceptibility, while another two columns illustrate temperature dependencies of the susceptibility at moderate value of the magnetic field. Let us at first make a few comments on typical thermal variations of the initial (zero-field) susceptibility times temperature product. It is evident from Fig.~\ref{fig:8} that the transverse component $\chi_x T$ generally shows monotonous decline upon decreasing temperature until it completely vanishes at zero temperature. The only exception to this rule is the particular case with the canting angle $2\alpha=90^{\circ}$, for which the product $\chi_x T$ initially reaches a round maximum and then it tends to some finite value as temperature goes to zero. On the other hand, the longitudinal component $\chi_z T$ mostly displays a monotonous increase upon decreasing temperature, which is followed by a divergence when approaching zero temperature. It should be pointed out that one may also detect for sufficiently large values of the tilting angle $\theta$ and canting angle $2\alpha$ also more striking thermal dependencies of the product $\chi_z T$, which exhibits upon decreasing temperature a round minimum successively followed by a zero-temperature divergence. The special case with the canting angle $2\alpha=90^{\circ}$ is again exceptional, since the product $\chi_z T$ monotonically decreases upon decreasing temperature until it reaches a constant value at zero temperature.

Furthermore, let us emphasize the most important consequences of the finite magnetic field on temperature dependencies of the susceptibility times temperature product. Apparently, the transverse component $\chi_x T$ diminishes upon decreasing temperature until it completely disappears at zero temperature independently of the tilting angle $\theta$ and the canting angle $2\alpha$. It could be thus concluded that the magnetic field does not significantly influence temperature variations of the transverse component $\chi_x T$ except the special case with the canting angle $2\alpha=90^{\circ}$.
By contrast, the magnetic field basically changes temperature variations of the longitudinal susceptibility times temperature product. It actually turns out that the longitudinal component $\chi_z T$ shows a monotonous decline upon decreasing temperature until it vanishes at zero temperature.

\begin{figure}[!th]
\includegraphics[scale=0.29]{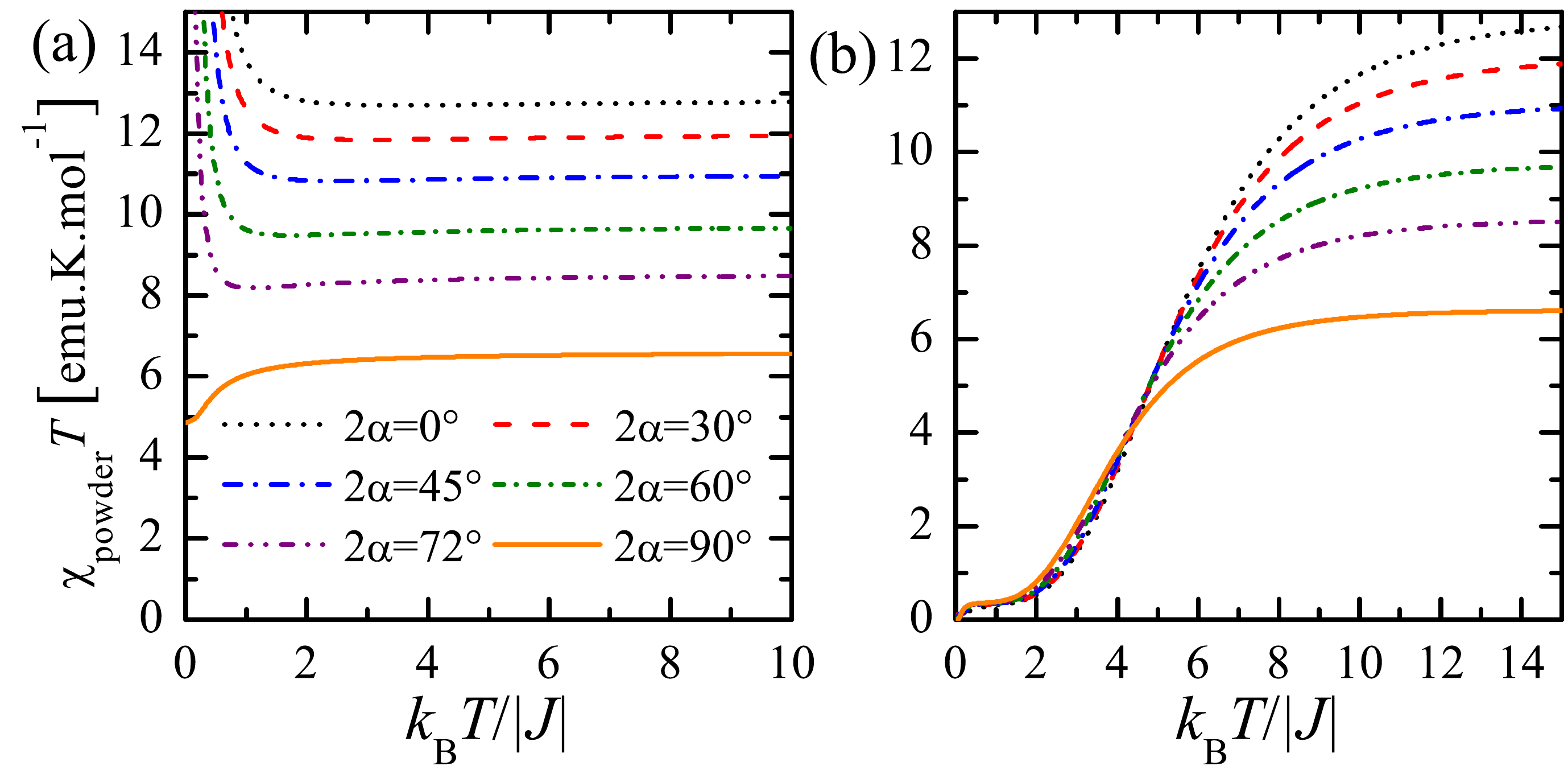}
\caption{\label{fig:10} (Color online) The powder susceptibility times temperature product $\chi_{powder} T$ as a function of dimensionless temperature $k_{\rm B} T/|J|$ for the fixed values of g-factors $g_{1}^{z_{1}}=g_{1}^{z_{2}}=20$, $g_{2}^{z}=g_{2}^{x}=2$ and two different magnetic fields: (a) $\mu_{\rm B} B/|J| = 0.0$; (b) $\mu_{\rm B} B/|J|=0.5$.}
\end{figure}

We will end up our discussion concerning with a typical thermal behavior of the susceptibility times temperature product by a comprehensive analysis of the powder susceptibility calculated from the longitudinal and transverse components according to Eq.~\ref{eq:35}. It can be observed from Fig.~\ref{fig:10}(a) that the initial (zero-field) powder susceptibility times temperature product is kept constant over a relatively wide temperature range before it exhibits a pronounced low-temperature divergence as temperature drops to zero. It can be easily understood from a comparison of Figs.~\ref{fig:8} and \ref{fig:10}(a) that the marked divergence of the powder susceptibility stems from the relevant behavior of the longitudinal component. Another interesting observation is that the high-temperature asymptotic value of the powder susceptibility times temperature product is the smaller, the greater the canting angle $2\alpha$ between two co-planar anisotropy axes is. As far as the temperature change of the powder susceptibility at finite magnetic fields is concerned, the powder susceptibility times temperature product monotonically decreases upon lowering temperature until it completely disappears at zero temperature.

\section{Heterobimetallic Dy-Cu compound}
\label{expthe}

In this section, we will take advantage of the exact solution of the spin-1/2 Ising-Heisenberg chain composed of regularly alternating Ising and Heisenberg spins in order to provide a deeper understanding of magnetic properties of the heterobimetallic coordination polymer Dy-Cu \cite{cal08}. It is worthwhile to recall that the polymeric complex Dy-Cu involves as a magnetic backbone 1D chain of exchange-coupled Dy$^{3+}$ and Cu$^{2+}$ ions (see Fig. \ref{dycu}), whereas the magnetic behavior of Dy$^{3+}$ and Cu$^{2+}$ ions can be reasonably well approximated at low enough temperatures by the notion of Ising and Heisenberg spins, respectively. The Ising nature of Dy$^{3+}$ magnetic ion can be attributed to a strong crystal-field splitting of the ground-state multiplet $^{6}$H$_{15/2}$ (f$^9$) with the total angular momentum $J_T=15/2$ ($L=5$, $S=5/2$) and the associated Land\'e g-factor $g = 4/3$ into eight well separated Kramers dublets. The Kramers dublet with the largest total angular momentum $J_T^z = \pm 15/2$ quantized with respect to the $z$-axis is several tens of Kelvins lower in energy compared to other Kramers dublets, which means that Dy$^{3+}$ magnetic ion can be effectively treated at sufficiently low temperatures as the two-valued Ising spin $\sigma = 1/2$ with the associated Land\'e g-factor $g = 2 (4/3) (15/2) = 20$. In this regard, the polymeric compound Dy-Cu affords the sought after experimental realization of the spin-$1/2$ Ising-Heisenberg chain with regularly alternating Ising and Heisenberg spins. It is also quite evident from Fig. \ref{dycu} that two crystallographically inequivalent orientations of coordination polyhedra of Dy$^{3+}$ ions exist in the Dy-Cu compound what might presumably imply two different local anisotropy axes of the Ising spins. It should be mentioned that the canting of local anisotropy axes has been disregarded in all foregoing studies \cite{str12,han13,roj14} and is the main subject matter of the present report.

\begin{figure}[!t]
\includegraphics[scale=0.29]{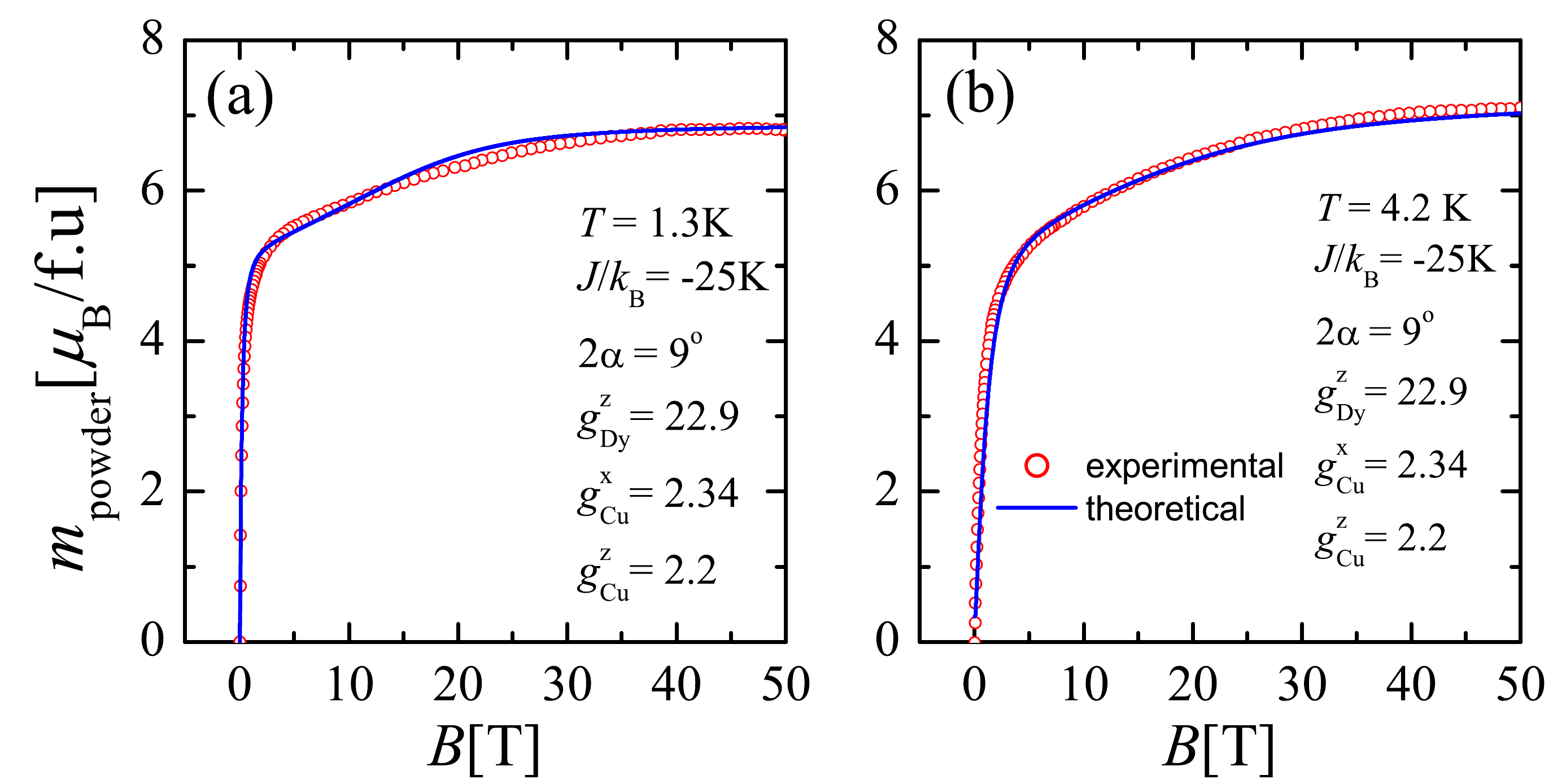}
\caption{\label{fig:12} (Color online) The high-field magnetization data recorded for the powder sample of Dy-Cu compound (red circles) at two different temperatures $T=1.3$K and $4.2$K versus the best theoretical fit (blue solid line) obtained with the help of the spin-1/2 Ising-Heisenberg chain (\ref{eq:1}) for the following set of fitting parameters $J/k_{\rm B} = -25$K, $g_{\rm Dy}^z \equiv g_1^{z_1} = g_1^{z_2} = 22.9$, $g_{\rm Cu}^x \equiv g_2^{x} = 2.34$, $g_{\rm Cu}^z \equiv g_2^{z} = 2.20$ and $2\alpha = 9^{\circ}$.}
\end{figure}

The high-field magnetization data of the powder sample of Dy-Cu compound are confronted in Fig.~\ref{fig:12}, after the relevant correction for the temperature-independent (Van Vleck) paramagnetism was made according to Ref. \cite{str12}, with the theoretical magnetization curve of the spin-1/2 Ising-Heisenberg chain (\ref{eq:1}) composed of regularly alternating Ising and Heisenberg spins. The best theoretical fit indicates an antiferromagnetic nature of the nearest-neighbor coupling between Dy$^{3+}$ and Cu$^{2+}$ ions, which makes from the polymeric compound Dy-Cu an intriguing experimental realization of a classical-quantum ferrimagnetic spin chain. The actual value of the exchange coupling between Dy$^{3+}$ and Cu$^{2+}$ ions $J/k_{\rm B} = -1.7$~K is however much smaller than the reported one $J/k_{\rm B} = -25$~K, because this latter value should be scaled down by the factor 15 as the true value of the total angular momentum of Dy$^{3+}$ ion is $J_T=15/2$ rather than $\sigma = 1/2$. The estimated values of g-factors are also adequately close to the typical values $g_{\rm Dy} \approx 20$ and $g_{\rm Cu} \approx 2.2$ for Dy$^{3+}$ and Cu$^{2+}$ ions. It is noteworthy that the theoretical value of g-factor $g_{\rm Dy} = 20$ would coincide with the actual magnetic moment of Dy$^{3+}$ ion with the total angular momentum $J_T = 15/2$ and g-value $g=4/3$ as far as it is effectively treated as the Ising spin $\sigma = 1/2$.

\begin{figure}[!t]
\includegraphics[scale=0.25]{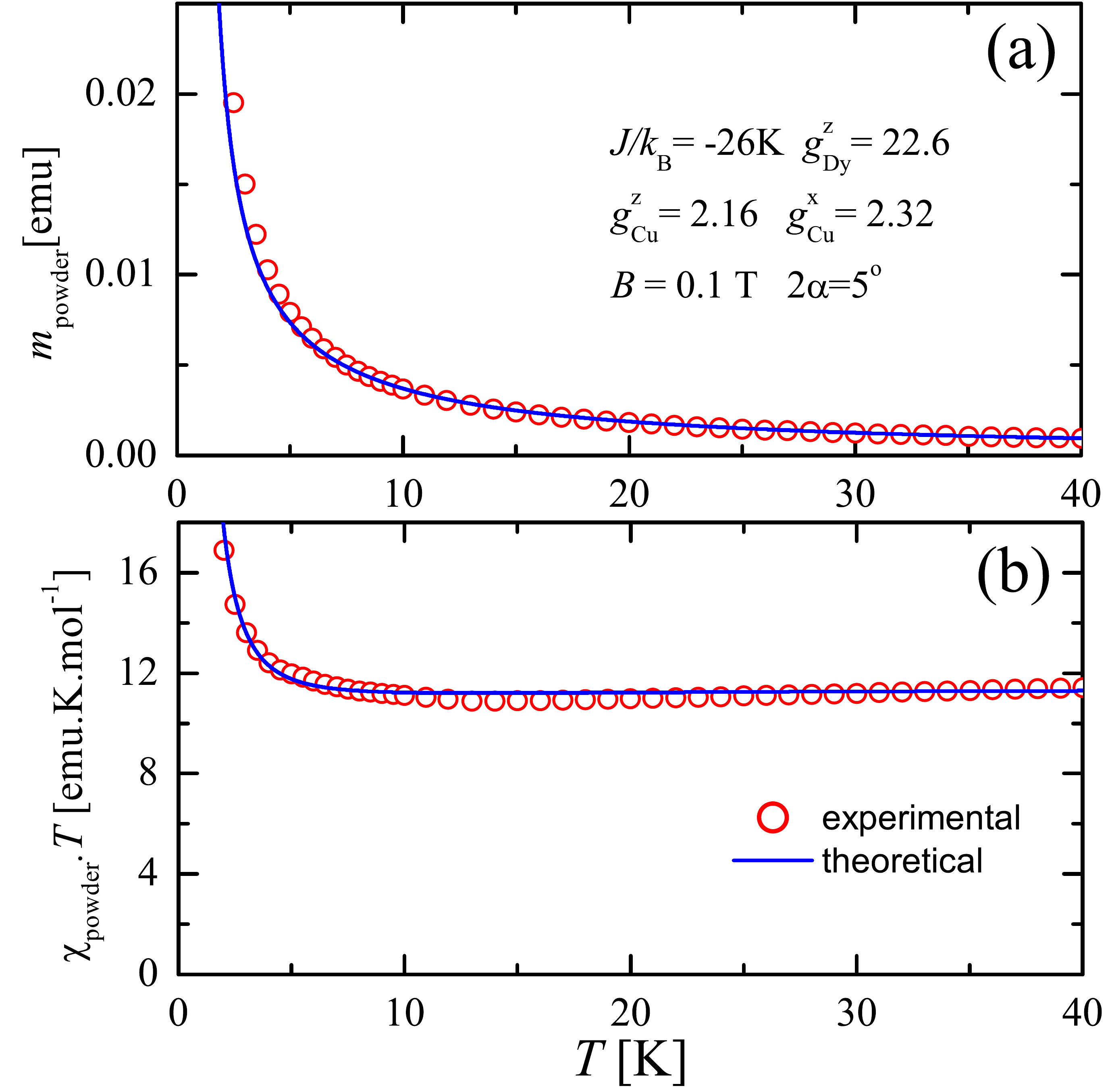}
\caption{\label{fig:13} (Color online) The temperature dependence of the magnetization and susceptibility times temperature data for the powder sample of Dy-Cu compound (red circles) at small magnetic field $B = 0.1$T along with the best theoretical fit (blue solid line) obtained with the help of the spin-1/2 Ising-Heisenberg chain (\ref{eq:1}) for the following fitting set: $J/k_{\rm B} = -26$K, $g_{\rm Dy}^z \equiv g_1^{z_1} = g_1^{z_2} = 22.6$, $g_{\rm Cu}^x \equiv g_2^{x} = 2.32$, $g_{\rm Cu}^z \equiv g_2^{z} = 2.16$ and $2\alpha = 5^{\circ}$.}
\end{figure}

Next, the low-field magnetization and susceptibility times temperature data for the powder sample of Dy-Cu compound are plotted in Fig.~\ref{fig:13} against temperature together with the relevant theoretical prediction obtained with the help of the spin-1/2 Ising-Heisenberg chain (\ref{eq:1}). The magnetization shows a relatively rapid decrease upon increasing temperature, whereas the relevant theoretical fit is in a reasonable accordance with the respective experimental data for the fitting set of parameters that nearly coincides with the one reported for the high-field magnetization curve. The susceptibility times temperature product exhibits a similar rapid decrease upon increasing temperature until it reaches almost constant value above $10$~K. Note furthermore that a plausible theoretical fit of the susceptibility times temperature data was obtained for the same set of fitting parameters as for the thermal variation of the magnetization. It should be emphasized, however, that the spin-1/2 Ising-Heisenberg chain of regularly alternating Ising and Heisenberg spins is applicable to Dy-Cu compound only in the low-temperature range $T<40$K, where the lowest-lying Kramers doublet of Dy$^{3+}$ ion is populated and correspondence with the Ising spins holds.

\section{Conclusion}
\label{conclusion}

In the present article we have comprehensively examined the ground-state phase diagram, magnetization process and susceptibility data of the spin-$1/2$ Ising-Heisenberg chain of regularly alternating Ising and Heisenberg spins in dependence on the canting angle between two different local anisotropy axes. It has been demonstrated that the investigated classical-quantum spin chain exhibits in the ground-state phase diagram two canted ferromagnetic and two canted antiferromagnetic phases. Another interesting finding concerns with the existence of a few macroscopically degenerate points, at which a perfect order of the Ising spins accompanies a complete disorder of the Heisenberg spins within the so-called 'half ice, half fire' ground states. Owing to this fact, the spin-$1/2$ Ising-Heisenberg chain unveils a remarkable diversity of zero-temperature magnetization curves, which may include true magnetization plateaus and jumps, quasi-plateaus or even smooth dependence of the magnetization on the external magnetic field depending on a relative orientation of the applied magnetic field with respect to the local anisotropy axes. Moreover, it has been verified that the averaging over all spatial orientations of the magnetic field generally causes smoothing of zero-temperature magnetization curves of the powder samples due to local quantum fluctuations, which mimic in certain respects the effect of thermal fluctuations reinforced by the rising temperature.

It has been convincingly evidenced that the investigated spin-$1/2$ Ising-Heisenberg chain of regularly alternating Ising and Heisenberg spins elucidates a low-temperature magnetic behavior of the coordination polymer Dy-Cu, which affords an outstanding experimental realization of a ferrimagnetic classical-quantum spin chain. It would be therefore highly desirable to probe angular dependence of the low-temperature magnetization process on a single-crystal sample of the coordination compound Dy-Cu, which is unfortunately not yet available. The preparation of the single-crystal sample of the polymeric compound Dy-Cu thus represents challenging task in view of the experimental testing of the most interesting findings presented in this work.

\appendix

\section{Residual entropy}

\begin{figure}
\includegraphics[scale=0.64]{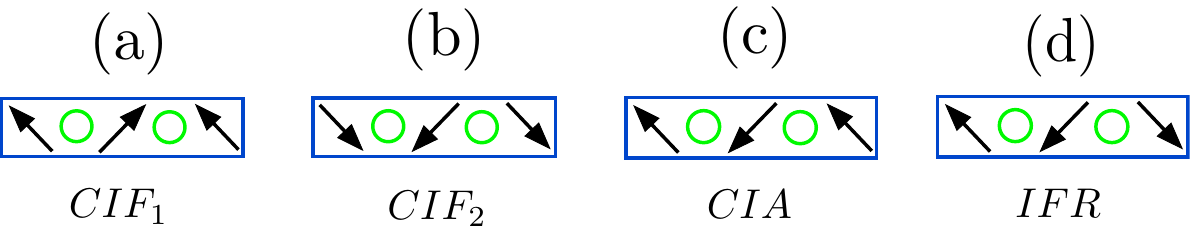}
\caption{\label{fig:14} (Color online) The spin configurations of unit cell with equal energy at the special coexistence point of the phases CIF$_1$, CIF$_2$ and CIA$_1$: (a) the spin configuration corresponding to the canted ferromagnetic phase CIF$_{1}$; (b) the spin configuration corresponding to the canted ferromagnetic phase CIF$_{2}$; (c) the spin configuration corresponding to the canted antiferromagnetic phase CIA$_{1}$; (d) the spin configuration IFR with the equal energy. Only the spin states of the Ising spins are visualized, whereas the Heisenberg spins denoted by circles are yalways in a quantum superposition given by $|\psi_{1}^{-}\rangle$ and $|\psi_{2}^{-}\rangle$.}
\end{figure}

The three phases CIF$_1$, CIF$_2$ and CIA$_1$ coexist together within a special frustration point of the spin-1/2 Ising-Heisenberg chain, where four  spin configurations of unit cell displayed in Fig.~\ref{fig:14} have equal energies. The spin configuration (a) corresponds to the canted ferromagnetic phase CIF$_{1}$, whereas the overall number of unit cells of this type is denoted as $N_{a}$. The spin configuration (b) corresponds to the canted ferromagnetic phase CIF$_{2}$, whereas the overall number of unit cells of this type is denoted as $N_{b}$. The spin configuration (c) corresponds to the canted antiferromagnetic phase CIA$_1$, whereas the overall number of unit cells of this type is denoted as $N_{c}$. Finally, the spin configuration (d) denoted as $IFR$ is the last spin configuration with the same energy,  configuration, whereas the overall number of unit cells of this type is denoted as $N_d$. It is worthy to mention that only the spin states of the Ising spins are relevant, because the Heisenberg spins denoted by circles are always in a quantum superposition given by $|\psi_{1}^{-}\rangle$ and $|\psi_{2}^{-}\rangle$.

The number of possible spin configurations of the entire chain with the fixed number of $N_a$, $N_b$, $N_c$, and $N_d$ cells is given by
\begin{equation}
\Omega=\frac{(N_{a}+N_{c})!}{N_{a}!N_{c}!}\frac{\left(\frac{N_{d}}{2}+N_{b}\right)!}{N_{b}!\left(\frac{N_{d}}{2}\right)!}
\frac{\left(N_{a}+N_{c}+\frac{N_{d}}{2}\right)!}{(N_{a}+N_{c})!\left(\frac{N_{d}}{2}\right)!},\label{eq:A1}
\end{equation}
where $N=N_{a}+N_{b}+N_{c}+N_{d}$ is the total number of unit cells. The first factor corresponds to all possible permutations between $N_a$ and $N_c$ cells. The second one stands for all possible permutations between $N_b$ cells and pairs of $N_d$ cells. The last factor accounts for the permutations of pairs of $N_d$ cells and $N_a$ or $N_c$ cells. Using the standard Lagrange multiplier technique, one can directly obtain the maximum number of states of the entire chain for the following particular choice of the numbers
\begin{alignat}{1}
N_{a}= & \frac{(5-\sqrt{5})}{10}N,\label{eq:A4}\\
N_{b}= & \frac{(10-4\sqrt{5})}{10}N,\label{eq:A5}\\
N_{c}= & \frac{(5-\sqrt{5})}{10}N,\label{eq:A6}\\
N_{d}= & \frac{(-10+6\sqrt{5})}{10}N.\label{eq:A7}
\end{alignat}
This choice leads to the overall degeneracy $\Omega=\left(\frac{\sqrt{5}+3}{2}\right)^{N}$, which yields the following residual entropy per spin  at the triple coexistence point between the phases CIF$_1$, CIF$_2$ and CIA$_1$ (red circles in Fig.~\ref{fig:2}(c-e)) is given by:
\begin{equation}
S=\frac{1}{2N}\ln\left(\Omega\right)=\frac{1}{2}\ln\left(\frac{\sqrt{5}+3}{2}\right).
\end{equation}

\end{document}